%% file: icc2018_NEW.tex
\documentclass[conference]{IEEEtran}

\usepackage{etex}
\usepackage{amsmath}                        
\usepackage{booktabs}                       
\usepackage{graphicx}
\usepackage{amssymb, subfigure, stfloats, amssymb, amsfonts, rotating, acronym}
\usepackage{multirow}
\usepackage{relsize}
\usepackage{array}
\usepackage{tabularx}
\usepackage{tikz}
\usetikzlibrary{patterns}
\usetikzlibrary{arrows,positioning,shapes.geometric,spy}
\usepackage{filecontents}
\usepackage{color}
\usepackage{xcolor}
\usepackage{pgfplots}
\usepackage{cite}
\usepackage{adjustbox}
\usepackage{colortbl}
\usepackage{breakurl}
\usepackage{url}
\usepackage{lipsum}
\usepackage[linesnumbered,algoruled,boxed,lined]{algorithm2e}
\usepackage[normalem]{ulem}
\newtheorem{example}{Example}

\DeclareMathOperator*{\sgn}{sgn}

\newcommand{\fixme}[2]{\ifx&#2&{\leavevmode\color{red}#1}\else{\leavevmode\color{red}FIXME\{}#1{\leavevmode\color{red}\}}\footnote{{\leavevmode\color{red}#2}}\PackageWarning{Fixme}{#1: #2}\fi}

\newcommand{\newstuff}[2]{\ifx&#2&{\leavevmode\color{blue}#1}\else{\leavevmode\color{blue}NEWSTUFF\{}#1{\leavevmode\color{blue}\}}\footnote{{\leavevmode\color{blue}#2}}\PackageWarning{Newstuff}{#1: #2}\fi}

\newcommand{\furkan}[2]{\ifx&#2&{\leavevmode\color{magenta}#1}\else{\leavevmode\color{magenta}FIXME\{}#1{\leavevmode\color{magenta}\}}\footnote{{\leavevmode\color{magenta}#2}}\PackageWarning{Furkan}{#1: #2}\fi}

\definecolor{bblue}{HTML}{4F81BD}
\definecolor{rred}{HTML}{C0504D}
\definecolor{ggreen}{HTML}{9BBB59}
\definecolor{ppurple}{HTML}{9F4C7C}

\begin{document}

\title{Partitioned Successive-Cancellation Flip Decoding of Polar Codes\footnotemark}

\author{\IEEEauthorblockN{Furkan Ercan, Carlo Condo, Seyyed Ali Hashemi, Warren J. Gross}
\IEEEauthorblockA{Department of Electrical and Computer Engineering, McGill University, Montr\'eal, Qu\'ebec, Canada\\
Email: furkan.ercan@mail.mcgill.ca, carlo.condo@mcgill.ca, seyyed.hashemi@mail.mcgill.ca, warren.gross@mcgill.ca}}

\maketitle

\begin{abstract}
Polar codes are a class of channel capacity achieving codes that has been selected for the next generation of wireless communication standards. Successive-cancellation (SC) is the first proposed decoding algorithm, suffering from mediocre error-correction performance at moderate code lengths. In order to improve the error-correction performance of SC, two approaches are available: (i) SC-List decoding which keeps a list of candidates by running a number of SC decoders in parallel, thus increasing the implementation complexity, and (ii) SC-Flip decoding that relies on a single SC module, and keeps the computational complexity close to SC. In this work, we propose the partitioned SC-Flip (PSCF) decoding algorithm, which outperforms SC-Flip in terms of error-correction performance and average computational complexity, leading to higher throughput and reduced energy consumption per codeword. We also introduce a partitioning scheme that best suits our PSCF decoder. Simulation results show that at equivalent frame error rate, PSCF has up to $5 \times$ less computational complexity than the SC-Flip decoder. At equivalent average number of iterations, the error-correction performance of PSCF outperforms SC-Flip by up to $0.15$ dB at frame error rate of $10^{-3}$.
\end{abstract}

\IEEEpeerreviewmaketitle

\section{Introduction}\label{sec:intro}

\IEEEPARstart{P}{olar} codes, introduced by Ar{\i}kan in \cite{arikan09}, are a class of error-correcting codes that provably achieves channel capacity when the code length approaches infinity. They have been selected for the control channel of the enhanced mobile broadband (eMBB) scenario of the $5^{\rm th}$ generation wireless systems standards (5G) \cite{3GPP-5G}. The standardization procedure is currently ongoing for other communication scenarios, such as massive machine type communications (mMTC). The mMTC communication scenario sees a large number of devices connected to each other, targeting low power/energy consumption, and improved error-correction performance \cite{towards_mmtc}. Therefore, practical algorithms for polar codes that would meet these requirements must be addressed.

\footnotetext{This version of the manuscript corrects an error in the previous ArXiv version, as well as the published version in IEEE Xplore under the same title, which has the DOI:10.1109/ICC.2018.8422464. The corrections include all the simulations of SC-Flip-based and SC-Oracle decoders, along with associated comments in-text.}

Successive-cancellation (SC) decoding of polar codes was introduced in \cite{arikan09}: its error-correction performance approaches channel capacity at infinite code length, but it degrades significantly at moderate to short code lengths. SC-List decoding \cite{TalList} improves the error-correction performance of polar codes significantly, at the cost of higher decoding latency and implementation complexity \cite{ercan-allerton}. On the other hand, successive-cancellation flip (SC-Flip) decoding \cite{SCFlip14} keeps the computational complexity close to that of SC, while providing error-correction performance close to that of SC-List.


Prior results in the literature about SC-Flip decoding target the correction of a single wrong decision in SC decoding \cite{SCFlip14}. If more errors are to be corrected, the decoding complexity grows linearly with the order of erroneous decisions that are targeted \cite{SCFlip17-jour}. In this work, we propose the partitioned SCFlip (PSCF) decoding algorithm: it subdivides the codeword into partitions, on which SC-Flip is run. PSCF targets the correction of at least a single wrong decision, with lower computational complexity than SC-Flip. We also present a codeword partitioning scheme that best suits our PSCF decoder, and aims at maximizing the error-correction performance of PSCF, while keeping the average computational complexity of it as close to that of SC as possible.

The remainder of this work is organized as follows: in Section~\ref{sec:bg}, an overview of polar codes and their decoding algorithms is presented. In Section~\ref{sec:pscf}, the PSCF decoding algorithm is detailed, while Section~\ref{sec:partitioning} describes a partitioning scheme based on the error observations obtained from SC decoding. Section~\ref{sec:results} reports simulation results, and conclusions are drawn in Section~\ref{sec:concl}.

\section{Preliminaries}\label{sec:bg}
\subsection{Polar Codes}

A polar code $PC(N,K)$ is a linear block code that can achieve channel capacity via channel polarization, that splits $N = 2^n, n \in \mathbb Z^+$ channel utilizations into $K$ reliable ones and $N-K$ unreliable ones. The reliable channels are used to transmit the information bits, while the unreliable channels are frozen to a known value, usually zero, leading to a code rate $R = K/N$.

\begin{figure}
  \centering
  \input{pc-enc.tikz}
  \caption{Polar code encoding for $PC(8,5)$.}
  \label{fig:polarencode}
\end{figure}
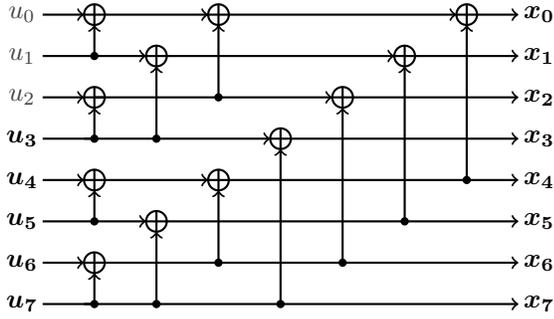

The encoding process of a polar code can be represented with the following matrix multiplication:
\begin{equation}\label{eqn:enc}
\boldsymbol{x_0^{N-1}} = \boldsymbol{u_0^{N-1}}G^{\otimes n}\text{,}
\end{equation}
where $\boldsymbol{x_0^{N-1}} = \{x_0,x_1,\ldots,x_{N-1}\}$ is the encoded vector, $\boldsymbol{u_0^{N-1}} = \{u_0,u_1,\ldots,u_{N-1}\}$ is the input vector, and the generator matrix $G^{\otimes n}$ is the $n$-th Kronecker product of the base polar code matrix $G = \left[\begin{smallmatrix} 1&0\\ 1&1 \end{smallmatrix} \right]$. Thus, a polar code of length $N$ can be seen as the concatenation of two polar codes of length $N/2$. 
The encoding operation in (\ref{eqn:enc}) for polar code $PC(8,5)$ is portrayed in Fig. \ref{fig:polarencode}; gray indices represent the frozen bits whereas the black indices indicate the information bits.

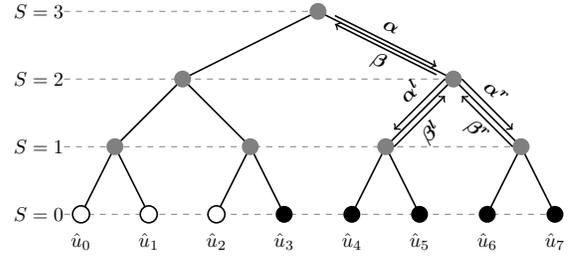
\begin{figure}
  \centering
   \scalebox{0.75}{\input{sc-dec.tikz}}
  \\
  \vspace{2pt}
  \caption{Successive-cancellation decoding tree for a $PC(8,5)$ code.}
  \label{fig:scdecode}
\end{figure}

The decoding process of the SC algorithm can be interpreted as a binary tree search: the tree is explored depth-first, with priority given to the left branch, with a complexity of $O(N\log N)$, and produces an estimated bit vector $\boldsymbol{\hat{u}_0^{N-1}}$.
The SC decoding tree for $PC(8,5)$ is depicted in Fig \ref{fig:scdecode}. Each parent node at stage $S$ contains logarithmic likelihood ratio (LLR) values $\boldsymbol{\alpha}=\{\alpha_0, \alpha_1, \ldots, \alpha_{2^S-1}\}$, which are passed to the child nodes via left and right operations recursively. From a parent node at stage $S$, the LLR values passed to left $\boldsymbol{\alpha^l} = \{\alpha^l_0, \alpha^l_1, \ldots, \alpha^l_{2^{S-1}-1}\}$ and right $\boldsymbol{\alpha^r} = \{\alpha^r_0, \alpha^r_1, \ldots, \alpha^r_{2^{S-1}-1}\}$ child nodes are approximated as
\begin{align}
{\alpha}^l_i &= \sgn(\alpha_{i})\sgn(\alpha_{i+2^{S-1}}) \min(|\alpha_{i}|,|\alpha_{i+2^{S-1}}|) \text{,} \label{eqn:alphaleft}\\
{\alpha}^r_i &= \alpha_{i+2^{S-1}} + (1-2\beta^{l}_{i})\alpha_{i} \text{.} \label{eqn:alpharight}
\end{align}
The LLRs at the root node are initialized with the channel LLR values $\boldsymbol{y}_0^{N-1}$. The partial sums $\boldsymbol{\beta}$ observed from the left $\boldsymbol{\beta^{l}}=\{\beta^l_0, \beta^l_1, \ldots, \beta^l_{2^{S-1}-1}\}$ and right $\boldsymbol{\beta^{r}}=\{\beta^r_0, \beta^r_1, \ldots, \beta^r_{2^{S-1}-1}\}$ child nodes are passed to their parent nodes as
\begin{equation}\label{eqn:beta}
  \beta_i=\left\{
  \begin{array}{@{}ll@{}}
    \beta^{l}_{i} \oplus \beta^{r}_{i}, & \text{if}~ i \leq 2^{S-1} \\
    \beta^{r}_{i}, & \text{otherwise.}
  \end{array}\right.
\end{equation}
where $\boldsymbol{\oplus}$ denotes bitwise XOR operation, and $0 \leq i < 2^S$. The $\beta$ value at leaf nodes is calculated as
\begin{equation}\label{eqn:bitestimate-sc}
\beta_{i}=\left\{
  \begin{array}{@{}ll@{}}
    0, & \text{when } \alpha_i \geq 0 \text{ } \text{or } i \in \Phi; \\
    1, & \text{otherwise.}
  \end{array}\right.
\end{equation}
where $\Phi$ denotes the set of frozen indices.

SC-List decoding algorithm \cite{TalList} creates $L$ distinct SC decoding paths working in parallel to have an improved error-correction performance. A path metric associated with each decoding path indicates the likelihood of the correct codeword. An outer cyclic-redundancy check (CRC) code improves the error-correction performance of SC-List decoding significantly. The computational complexity of the SC-List decoder is $O(LN\log N)$.

\subsection{Successive-Cancellation Flip Decoding}
In \cite{SCFlip14}, the SC-Flip decoding algorithm was introduced. It was observed that, when an SC decoder fails to estimate the correct codeword, it is either due to a wrong decision that is caused by the channel noise, or due to a prior wrong decision that was made earlier in the SC tree. It was also explained that the first wrong decision that occurs while decoding is always due to channel noise. Moreover, experiments show when SC decoding fails, it is mostly due to a single wrong decision caused by the channel which is potentially followed by propagated wrong decisions. A hypothetical decoder, called SC-Oracle decoder, was created to show that if all first wrong decisions are avoided, the error-correction performance of SC decoding would improve significantly.

The SC-Flip algorithm attempts to identify and correct the first error due to channel noise that the SC algorithm would incur. To do so, a CRC outer code with a remainder of $C$ bits is used to encode the information bits. At the end of a SC decoding phase, if the CRC does not detect any error, the estimated codeword is assumed to be correct. If not, a number of indices corresponding to low-reliability decisions are stored and sorted, then a second iteration is initiated. The bit associated to the index with the least reliable soft information is flipped, and SC is applied to the remainder of the decoding tree, followed by a CRC check. This process is repeated considering the stored low-reliability decision indices until either the CRC passes, or a maximum number of iterations $T_{max}$ is performed.

The computational complexity of SC-Flip decoding is $O(N \log N [1 + T_{max} \times \text{Pr}(R,\text{SNR})])$ \cite{SCFlip14}, where $\text{Pr}(R,\text{SNR})$ denotes the frame error rate (FER) of a polar code of rate $R$ at given signal-to-noise ratio (SNR) under SC decoding. Consequently, the average computational complexity of SC-Flip decoding is directly proportional to the average number of iterations, that depends on both $T_{max}$ and $\text{Pr}(R,\text{SNR})$. 

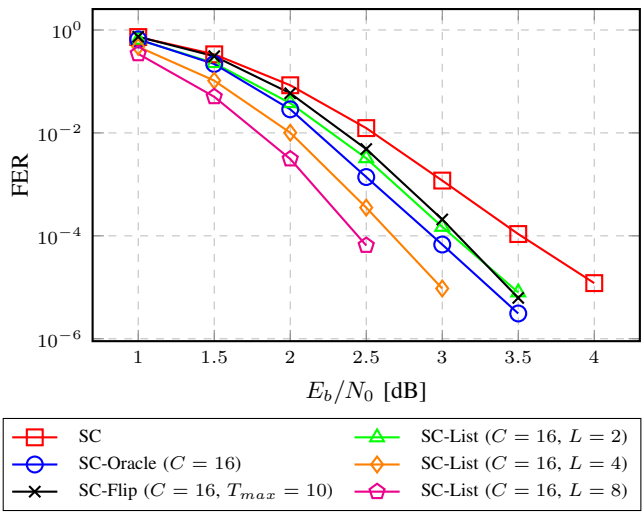
\begin{figure}
  \centering
\input{SCF-perf.tikz}
  \\
  \vspace{2pt}
  \ref{SCF-perf}
  \caption{SC-Flip decoding FER performance compared with SC and SC-Oracle with $PC(1024,512)$. $T_{max}=10$, and CRC length is $16$.}
  \label{fig:SCF-perf}
\end{figure}

Fig. \ref{fig:SCF-perf} presents the performance of SC-Flip compared to SC, SC-Oracle and SC-List for $PC(1024,512)$ constructed for a SNR of $2.5$ dB. 

It can be seen that while the error-correction performance of SC-Oracle lies in between of SC-List performances with $L=2$ and $L=4$, SC-Flip matches the FER of SC-List with $L=2$.
The performance gap between SC-Flip and SC-Oracle is due to two reasons: either the estimated codeword with a correct CRC check still contains errors, or the decoding stopped after reaching the maximum number of iterations without being successful. Note that adding CRC bits to a polar code affects its error-correction performance, as it is effectively adding more non-frozen bits.

Further improvements for SC-Flip decoding algorithm have been recently proposed in \cite{SCFlip17-conf} and \cite{SCFlip17-jour}: a generalized SC-Flip decoder algorithm uses nested flips to correct more than one erroneous decision with a single CRC. They also introduce a metric that helps the baseline SC-Flip decoder to detect the erroneous bit indices more accurately. Their simulation results show up to $0.8$~dB improvement over the error-correction performance of SC-Flip. On the other hand, their implementation requires an excessive number of iterations.

\section{Partitioned SC-Flip Decoding}\label{sec:pscf}
Let Pr$(E_i)$ denote the probability of failed decoding for an SC decoder, where $E_i$ represents the number of channel-induced errors with $0 < i \leq K+C$, and let Pr$(E_0)$ denote the probability of a successful decoding. Thus:
\begin{equation}\label{eqn:pr-sc}
\text{Pr}(E_0) + \text{Pr}(E_1) + \sum_{i = 2}^{K+C} \text{Pr}(E_i) = 1 
\end{equation}

SC-Flip attempts to minimize Pr$(E_1)$ within a maximum number of iterations, but it cannot help with Pr$(E_i)$ when $i >1$. The ability to detect and correct more than a single error in a codeword would improve the error-correction performance significantly. We propose a partitioned SC-Flip (PSCF) decoding algorithm, where the estimated codeword is divided into sub-blocks, with a partitioning factor $P$. Each partition is protected with its own CRC, all of which are independent from each other.
\begin{example}
To have a better understanding of how PSCF helps minimizing erroneous decisions, with a partitioning factor of $P=2$, the error probabilities $\text{Pr}(E_1)$ and $\text{Pr}(E_2)$ in (\ref{eqn:pr-sc}) can be reinterpreted as:
\begin{align}\label{eqn:pr-pscf-e1}
\text{Pr}(E_1) = & \text{Pr}(e_0 \in p_1) \times \text{Pr}(e_1 \in p_2) +\nonumber \\ 
				 & \text{Pr}(e_1 \in p_1) \times \text{Pr}(e_0 \in p_2)
\end{align}
and
\begin{align}\label{eqn:pr-pscf-e2}
\text{Pr}(E_2) = & \text{Pr}(e_0 \in p_1) \times \text{Pr}(e_2 \in p_2) +\nonumber \\ 
				 & \text{Pr}(e_1 \in p_1) \times \text{Pr}(e_1 \in p_2) + \\
				 & \text{Pr}(e_2 \in p_1) \times \text{Pr}(e_0 \in p_2)\nonumber
\end{align}
where $e_i \in p_j$ indicates that $i$ errors are present in partition $p_j$, with $0 \leq i \leq 2$ and $0 < j \leq 2$.

In SC-Flip, the CRC enables the algorithm to correct a single error. Dividing the codeword in two partitions, each protected by its own CRC, allows to correct up to one error in each partition, as expressed by the following error probability:
\begin{align}\label{eqn:pr-pscf-all}
& \text{Pr}(e_0 \in p_1) \times \text{Pr}(e_1 \in p_2) + \nonumber \\
& \text{Pr}(e_1 \in p_1) \times \text{Pr}(e_0 \in p_2) + \\
& \text{Pr}(e_1 \in p_1) \times \text{Pr}(e_1 \in p_2)  \nonumber
\end{align}

Note that if multiple channel errors occur in a single partition, a successful decoding is not possible with PSCF. As a result, PSCF can detect and correct more than one error if each error resides in a different partition.
\end{example}

The PSCF decoding process is described in Algorithm \ref{alg:pscf}. The information bit indices $\boldsymbol{I}$ and partitioning indices $\boldsymbol{\rho}$ are predetermined and known by the decoder. For each partition, the SC algorithm is executed first, followed by the computation of the CRC remainder (lines 4-5). If the CRC detects an error for the first time, then the indices of the $T_{max}$ information bits that have the least reliable LLRs are identified (line 7). For a maximum of $T_{max}$ iterations, the SC algorithm is executed: at each iteration $t$, the information bit with the $t^{th}$ least reliable LLR is flipped, until the CRC does not detect an error anymore (lines 9 to 12). If after $T_{max}$ iterations the CRC still detects an error, then the decoding process is terminated. 

\begin{algorithm}[t]
\SetKwData{Left}{left}\SetKwData{This}{this}\SetKwData{Up}{up}
\SetKwFunction{Union}{Union}\SetKwFunction{FindCompress}{FindCompress}
\SetKwInOut{Input}{input}\SetKwInOut{Output}{output}
  \Input{$\boldsymbol{y_{0}^{N-1}},T_{max},\boldsymbol{\rho_{1}^{P}},\boldsymbol{I}$}
  \Output{$\hat{u}_{0}^{N-1}$}
  $\rho[0] = 0$\\
  \For{$j = 1$ \KwTo $P$}{
  	\For{$i = \rho[j-1]$ \KwTo $\rho[j]$}{
  		$(\boldsymbol{\hat{u}_{\rho[j-1]}^{\rho[j]}}, \boldsymbol{\alpha_{\hat{u}_i}}) \leftarrow  \text{SC}(\boldsymbol{y_{0}^{N-1}},\boldsymbol{I},\O)$	\\
    	\If{$T_{max}>1 ~\text{\rm \bf and } \textsc{CRC}(\boldsymbol{\hat{u}_{\rho[j-1]}^{\rho[j]}}) \text{ \rm fails}$}{
    	$\boldsymbol{\alpha_{sort}} = sort(|\boldsymbol{\alpha_{\hat{u}_i}}|), i \in \boldsymbol{I}$\\
    	$\boldsymbol{U} = \text{first } T_{max} \text{ indices of } \boldsymbol{\alpha_{sort}}$\\
    	$t = 1$\\
			\While{$t \leq T_{max} ~\text{\rm \bf and }  \textsc{CRC}(\boldsymbol{\hat{u}_{\rho[j-1]}^{\rho[j]}}) ~\text{\rm fails} $}{			
			 
				$(\boldsymbol{\hat{u}_{\rho[j-1]}^{\rho[j]}}, \boldsymbol{\alpha_{\hat{u}_i}})  \leftarrow \text{SC}(\boldsymbol{y_{0}^{N-1}},\boldsymbol{I},U[t])$	\\	
					$t = t+1$
			}
			\If{$\textsc{CRC}(\boldsymbol{\hat{u}_{\rho[j-1]}^{\rho[j]}})\text{ \rm fails}$}{
			terminate process\\
			}
	}
  }
  }
\caption{Partitioned SC-Flip Algorithm}\label{alg:pscf}
\end{algorithm}

In order for the code rate to remain the same when either PSCF or SC-Flip are applied, the total number of information and CRC bits are unchanged, such that
\begin{align}
\sum_{i=1}^{P} K_{p_i} = K, \label{eqn:k-dist} \\
P \times C_{p_i} = C  \label{eqn:crc-dist}
\end{align}
where $K_{p_i}$ ($C_{p_i}$) is the total number of information (CRC) bits in partition $p_i$ of PSCF, and $K$ ($C$) is the total number of information (CRC) bits in SC-Flip. In order to keep the effective rate of PSCF equal to that of SC-Flip, the number of CRC bits in both cases have to be the same. The most straightforward method to keep the same effective rate is to distribute the CRC bits equally among the partitions as suggested in (\ref{eqn:crc-dist}). 

Depending on the number of partitions and their position, the number of information bits included in each partition might be different. After the information bits, each partition reserves the following $C_\text{SCF}/P$ most reliable position to the CRC remainder bits. As a result, the bits assigned to the CRC in SC-Flip and PSCF are different, while the locations of the information bits are the same under both algorithms.

Note that the idea of partitioning was also used in SC-List decoders in \cite{PSCL-GLOBECOM,PSCL-ICASSP,PSCL-JETCAS}: nevertheless, SC-List partitioning involves a completely different process, and aims at a different outcome. Partitioned SC-List (PSCL) divides the SC decoding tree in upper and lower tree, using a lower list size in the upper part to minimize memory requirements without degrading the error-correction performance. On the other hand, the PSCF algorithm loops SC-Flip over different portions of the codeword, reducing the average number of iterations and improving error-correction performance.

\section{Codeword Partitioning}\label{sec:partitioning}

Careful partitioning of the SC-Flip decoding process can significantly reduce the average number of performed iterations, and improve the error-correction performance. As mentioned in Section \ref{sec:pscf}, with partitioning factor $P$, PSCF can identify and correct up to $P$ errors. In this Section, we refer to an error pattern of $n^\text{th}$ order when $n$ errors occur in the codeword, and represent it with $E_n$. 

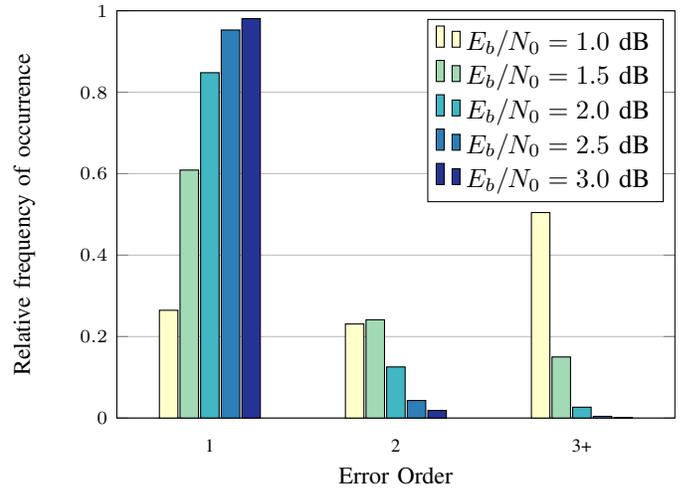
\begin{figure}
  \centering
\input{errordist2.tikz}
  \vspace{-6pt}
  \caption{Error order distribution for $PC(1024,512)$ under SC decoding with various $E_b/N_0$ points.}
  \label{fig:errordist}
\end{figure}

Fig. \ref{fig:errordist} depicts the distribution of errors according to their order, for $PC(1024,512)$, under SC decoding. The probability of a failed decoding being due to $E_1$ increases with $E_b/N_0$. 
For example, at $E_b/N_0 = 2.5$ dB, $95.3\%$ of decoding failures are due to $E_1$. Therefore, at high $E_b/N_0$, the ability to correct error orders higher than $E_1$ becomes an advantage for PSCF only when the proposed algorithm is as effective as SC-Flip in correcting failures due to $E_1$.

In general, the ability to identify and correct errors in the codeword improves with the CRC size. As mentioned in Section \ref{sec:pscf}, the CRC bits for each partition are uniformly distributed over partitions for PSCF decoding. As a result, each partition should cover an equal probability of error occurrences. Since $E_1$ dominates the probability of error occurrence at medium and high $E_b/N_0$ values, we can approximate an equal error probability partitioning method by dividing the codeword with respect to $E_1$. In order to distribute the partitions, given the length and rate of a polar code, a map of error distribution is required. 

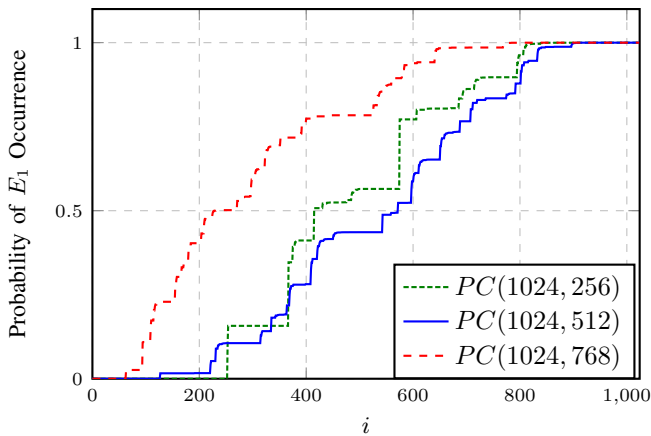
\begin{figure}
  \centering
\input{CDFerrordist.tikz}
  \vspace{-6pt}
  \caption{Cumulative error distribution for $E_1$ for polar codes of length $N=1024$ and rates $R \in \{ 1/4,1/2,3/4 \}$ at $E_b/N_0 = 2.5$ dB.}
  \label{fig:CDFerrordist}
\end{figure}

Fig. \ref{fig:CDFerrordist} portrays the cumulative probability of $E_1$ occurrence over a set of polar codes where $N=1024$ and rates $R \in \{ \frac{1}{4},\frac{1}{2},\frac{3}{4} \}$, under SC decoding, at $E_b/N_0 =  2.5$ dB. These curves have been obtained through SC-Oracle decoding, storing the error indices for failed decoding due to a single error. The partitioning indices $\boldsymbol{\rho}$ should be placed according to $E_1$ distribution, given that each partition should cover $1/P$ of the $E_1$ errors. Note that the partitioning indices refer to the last bits of each partition. It can be seen in Fig. \ref{fig:CDFerrordist} that the partitioning index $\boldsymbol{\rho}$ does not only change with $P$ but also with rate $R$. For example, for a partitioning factor of $P=2$, the first partitioning index $\rho_1$ for $PC(1024,512)$ should correspond to first the $50\%$ of $E_1$ distribution and thus should be located around $N/2$, while for $PC(1024,768)$ the $50\%$ mark is reached at $\approx N/5$.

If a consecutive series of bits have zero error probability, they are represented by a horizontal flat line in the corresponding cumulative error distribution. These bits correspond to either frozen channels or extremely reliable information channels. If a partitioning index corresponds to such a flat line, $\rho_i$ can be placed anywhere within the set of bits without affecting the error-correction performance of PSCF. If $\rho_i$ is placed at the highest bit index in the set, when the partition is reiterated, operations for the complete set have to be repeated. On the other hand, if $\rho_i$ is placed at the lowest bit index in the set, the flat line is included at the beginning of the following partition, and never reiterated. Such an example can be found for $PC(1024,512)$, where the $45\%$ mark of $E_1$ corresponds to a flat line for $P=2$ in Fig. \ref{fig:CDFerrordist}.

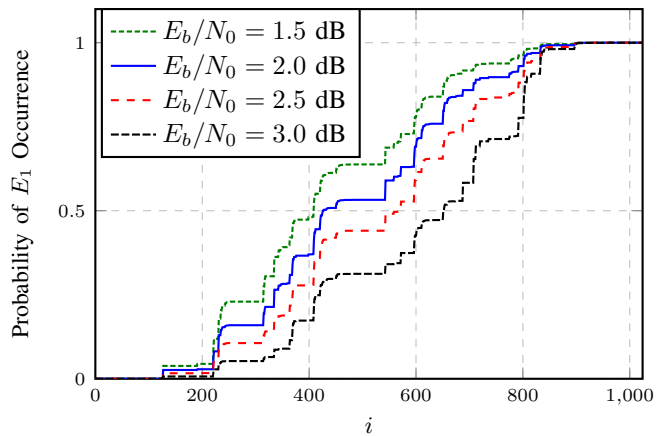
\begin{figure}
  \centering
\input{CDFerrordist-ebn0.tikz}
  \vspace{-6pt}
  \caption{Cumulative error distribution for $E_1$ for $PC(1024,512)$ at various $E_b/N_0$ points.}
  \label{fig:CDFerrordist-ebn0}
\end{figure}

Fig. \ref{fig:CDFerrordist-ebn0} shows how the cumulative $E_1$ distribution changes within a set of $E_b/N_0$ values for $PC(1024,512)$. It can be seen that the error distribution does not only depend on the rate but also on signal-to-noise ratio. With increasing $E_b/N_0$, the likelihood of observing an error in the codeword decreases: when errors indeed occur, they are more likely to happen among the least reliable of the information bits. The information and frozen bit indices used to obtain the curves in Fig. \ref{fig:CDFerrordist-ebn0} are the same for all cases, and are optimized for $E_b/N_0=2.5$ dB. That means that when $E_b/N_0\ne 2.5$ dB, bits considered reliable can be less or more so, and vice versa. This phenomenon explains the shift towards the right of the cumulative $E_1$ distribution as $E_b/N_0$ increases.  It can be noticed that some $E_1$ increments are more substantial at higher $E_b/N_0$: these are relative to bit indices $i$ associated with some of the least reliable information bits (e.g $i=708$ and $i=802$). As the channel conditions improve, these indices refer to more and more unreliable bits, thus leading to an increased probability of $E_1$ occurring at those indices. At the same time, other bit indices among the  least reliable information bits (e.g. $i=221$), improve their reliability as $E_b/N_0$ increases, leading to lower and lower probability of $E_1$ occurring at that position.

\section{Simulation Results}\label{sec:results}

\begin{figure}
  \centering
\input{PSCF-iter.tikz}
  \vspace{2pt}
  \caption{Normalized average computational complexity of PSCF($P \in \{2,4\}$) and SC-Flip for $PC(1024,512)$ at matched error-correction performance ($C=16$, $T_{max}=10$).}
  \label{fig:PSCF-iter}
\end{figure}
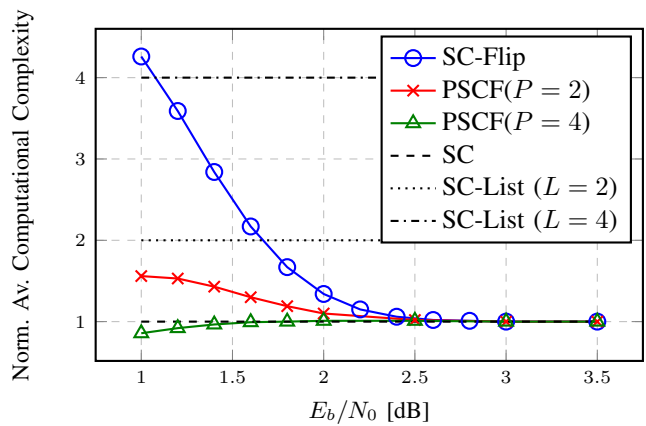

As mentioned in Section \ref{sec:bg}, the average computational complexity of SC-Flip decoder, and thus its latency, is directly proportional to the average number of iterations. In this context, the computational complexity of PSCF is also related to its average number of iterations. Fig. \ref{fig:PSCF-iter} compares the normalized average computational complexity of PSCF ($P \in \{2,4\}$) with SC-Flip for $PC(1024,512)$. $T_{max}=10$ for SC-Flip. Comparisons are made at equivalent FER. We consider the original SC-Flip algorithm as the baseline comparison for PSCF: the improvements proposed in \cite{SCFlip17-conf,SCFlip17-jour} can be applied to both, independently.
At low $E_b/N_0$, the average computational complexity of SC-Flip is as high as that of SC-List decoding with list size of $L=4$. On the other hand, the worst case computational complexity of PSCF with $P=2$ is only $55\%$ above that of SC. At low $E_b/N_0$ values, the complexity of PSCF with $P=4$ is less than that of SC: this is due to the early termination of decoding in case a partition fails after $T_{max}$ iterations. At higher $E_b/N_0$, it converges to the complexity of SC. From Fig. \ref{fig:PSCF-iter} it can be seen that, compared to SC-Flip, PSCF is up to $2.7\times$ faster with $P=2$, and up to $5\times$ faster with $P=4$.

\begin{figure}[t]
  \centering
\input{PSCF-perf-P2.tikz}
  \vspace{2pt}
  \ref{PSCF-perf}
  \caption{FER performance comparison of PSCF with $P = 2$ against SC-Flip at matched iterations for $E_b/N_0 \in \{1.0,1.5\}$ dB, also compared with SC-List ($L=2$) and SC-Oracle for $PC(1024,512)$. ($C=16$, $T_{max}=10$)}
  \label{fig:PSCF-perf-P2}
\end{figure}
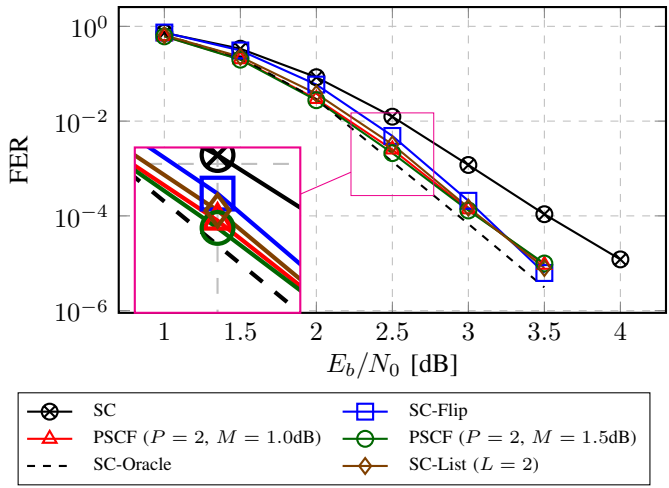

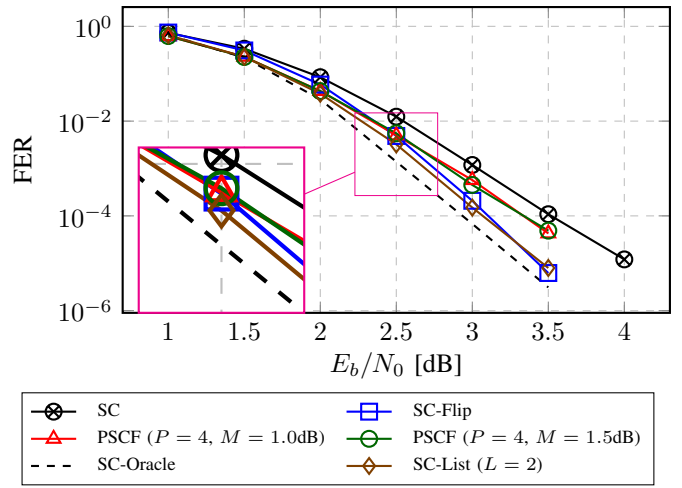
\begin{figure}[t]
  \centering
\input{PSCF-perf-P4.tikz}
  \vspace{2pt}
  \ref{PSCF-perf}
  \caption{FER performance comparison of PSCF with $P = 4$ against SC-Flip at matched iterations for $E_b/N_0 \in \{1.0,1.5\}$ dB, also compared with SC-List ($L=2$) and SC-Oracle for $PC(1024,512)$. ($C=16$, $T_{max}=10$)}
  \label{fig:PSCF-perf-P4}
\end{figure}

The error-correction performance of PSCF with different numbers of partitions is depicted in Fig. \ref{fig:PSCF-perf-P2} and Fig. \ref{fig:PSCF-perf-P4}. The curves have been obtained by matching the average number of iterations at the $E_b/N_0$ point $M$. The partitioning indices $\boldsymbol{\rho}$ are selected based on the partitioning scheme described in Section \ref{sec:partitioning}. It can be observed that at low $E_b/N_0$ points, PSCF outperforms SC-Oracle, as it can correct more than a single error. The advantage of PSCF over SC-Oracle reduces as $E_b/N_0$ increases. This can be explained observing Fig. \ref{fig:errordist}, where the probability of a single error causing a failed decoding increases with respect to higher error orders.

Compared to SC-Flip, PSCF has better error-correction performance in most cases. At low $E_b/N_0$, PSCF has a lower FER because of its ability to correct higher-order errors. As $E_b/N_0$ grows, the impact of higher-order error correction begins to decrease; however, PSCF performs better than SC-Flip in terms of correcting single errors. This is due to the fact that PSCF has, overall, the ability to flip up to $P\times T_{max}$ bits, increasing the probability of identifying the wrong decision.
As $E_b/N_0$ grows further, SC-Flip begins to gain advantage over PSCF. The reason is that at high $E_b/N_0$ values, as mentioned in Section \ref{sec:pscf}, the sub-optimal CRC placement of PSCF due to partitioning makes PSCF more vulnerable to errors than SC-Flip. With increasing partitioning factor $P$, CRC distribution gets more sub-optimal (see Fig. \ref{fig:PSCF-perf-P4}). Nevertheless, the performance of PSCF with $P=2$ is better than SC-Flip at practical FER region of $[10^{-3};10^{-4}]$. With $C=16$ bits for $PC(1024,512)$, PSCF algorithm outperforms SC-Flip by up to $0.15$ dB with $P=2$ in the target FER region. Finally, since PSCF with $P=2$ performs closer to SC-Oracle than SC-Flip, its performance is the closest to SC-List with $L=4$.

\section{Conclusion}\label{sec:concl}
In this work, we present the partitioned successive-cancellation flip (PSCF) decoding algorithm, that divides the polar code decoding tree into $P$ partitions, and applies the successive-cancellation flip (SC-Flip) algorithm to each partition separately. We show that with partitioning, unlike with SC-Flip, it is possible to correct more than one erroneous bit estimation, as long as the wrong decisions take place in separate partitions. We also show that the average number of iterations can be reduced significantly with partitioning. Then, we present a partitioning scheme for PSCF based on the probability of error distribution for a given codeword. At equivalent number of iterations, our approach demonstrates an improved error-correction performance of up to $0.15$ dB with a partitioning factor of $P=2$ compared to SC-Flip decoding. At equivalent error-correction performance, PSCF shows an average computational complexity reduction of $2.7\times$ with $P=2$, and of $5\times$ with $P=4$ compared to SC-Flip. In case of $P=4$, the overall average computational complexity is equivalent to that of a single SC decoder. This leads to increased average throughput and reduced energy consumption for the PSCF decoder.

\end{document}

%% file: pc-enc.tikz
\begin{tikzpicture}[scale=.55, thick]
  \node [color=darkgray] at (.5,0) {$u_0$} ;
  \node [color=darkgray]at (.5,-1) {$u_1$};
  \node [color=darkgray]at (.5,-2) {$u_2$};
  \node at (.5,-3) {$\boldsymbol{u_3}$};
  \node at (.5,-4) {$\boldsymbol{u_4}$};
  \node at (.5,-5) {$\boldsymbol{u_5}$};
  \node at (.5,-6) {$\boldsymbol{u_6}$};
  \node at (.5,-7) {$\boldsymbol{u_7}$};

  \foreach \x in {-6,-4,-2,0}
  {
    \draw [->] (1,\x) -- (2,\x);
    \draw (1,\x-1) -- (2.25,\x-1);

    \draw (2.25,\x) circle [radius=.25];
    \draw (2,\x) -- (2.5,\x);
    \draw (2.25,\x-.25) -- (2.25,\x+.25);

    \draw [->] (2.25,\x-1) -- (2.25,\x-.25);

    \fill (2.25,\x-1) circle [radius=.1];
  }

  \foreach \x in {-4,0}
  {
    \draw [->] (2.5,\x) -- (5,\x);
    \draw [->] (2.25,\x-1) -- (3.5,\x-1);

    \draw (5.25,\x) circle [radius=.25];
    \draw (5,\x) -- (5.5,\x);
    \draw (5.25,\x-.25) -- (5.25,\x+.25);

    \draw (3.75,\x-1) circle [radius=.25];
    \draw (3.5,\x-1) -- (4,\x-1);
    \draw (3.75,\x-1-.25) -- (3.75,\x-1+.25);

    \draw [->] (2.25,\x-2) -- (5.25,\x-2) -- (5.25,\x-.25);
    \fill (5.25,\x-2) circle [radius=.1];
    \draw [->] (2,\x-3) -- (3.75,\x-3) -- (3.75,\x-1-.25);
    \fill (3.75,\x-3) circle [radius=.1];
  }

  \draw [->] (5.5,0) -- (11,0);
  \draw [->] (4,-1) -- (9.5,-1);
  \draw [->] (5.25,-2) -- (8,-2);
  \draw [->] (3.75,-3) -- (6.5,-3);

  \foreach \x in {-1,0}
  {
    \draw [->] (5.5+1.5*\x,\x-4) -- (11.25+1.5*\x,\x-4) -- (11.25+1.5*\x,\x-.25);
    \draw [->] (5.25+1.5*\x,\x-6) -- (11.25+1.5*\x-3,\x-6) -- (11.25+1.5*\x-3,\x-2-.25);
  }

  \foreach \x in {-3,...,0}
  {
    \draw (11.25+1.5*\x,\x) circle [radius=.25];
    \draw (11+1.5*\x,\x) -- (11.5+1.5*\x,\x);
    \draw (11.25+1.5*\x,\x-.25) -- (11.25+1.5*\x,\x+.25);

    \fill (11.25+1.5*\x,\x-4) circle [radius=.1];

    \draw [->] (11.5+1.5*\x,\x) -- (12.5,\x);
    \draw [->] (11.25+1.5*\x,\x-4) -- (12.5,\x-4);
  }

  \node at (13,0) {$\boldsymbol{x_0}$};
  \node at (13,-1) {$\boldsymbol{x_1}$};
  \node at (13,-2) {$\boldsymbol{x_2}$};
  \node at (13,-3) {$\boldsymbol{x_3}$};
  \node at (13,-4) {$\boldsymbol{x_4}$};
  \node at (13,-5) {$\boldsymbol{x_5}$};
  \node at (13,-6) {$\boldsymbol{x_6}$};
  \node at (13,-7) {$\boldsymbol{x_7}$};

\end{tikzpicture}

%% file: sc-dec.tikz
\begin{tikzpicture}[scale=.6, thick]

\draw [thin,gray,dashed] (0,-1) -- (7.5,-1);
\draw [thin,gray,dashed] (0,-3) -- (11.5,-3);
\draw [thin,gray,dashed] (0,-5) -- (13.5,-5);
\draw [thin,gray,dashed] (0,-7) -- (14.5,-7);

\node at (-.75,-1) {$S=3$};
\node at (-.75,-3) {$S=2$};
\node at (-.75,-5) {$S=1$};
\node at (-.75,-7) {$S=0$};

\draw (7.5,-1) -- (3.5,-3);
\draw (7.5,-1) -- (11.5,-3);

\draw (3.5,-3) -- (1.5,-5);
\draw (3.5,-3) -- (5.5,-5);
\draw (11.5,-3) -- (9.5,-5);
\draw (11.5,-3) -- (13.5,-5);

\draw (1.5,-5) -- (0.5,-7);
\draw (1.5,-5) -- (2.5,-7);

\draw (5.5,-5) -- (4.5,-7);
\draw (5.5,-5) -- (6.5,-7);

\draw (9.5,-5) -- (8.5,-7);
\draw (9.5,-5) -- (10.5,-7);

\draw (13.5,-5) -- (12.5,-7);
\draw (13.5,-5) -- (14.5,-7);


  \draw[black,fill=white] (.5,-7) circle [radius=.25];			
  \fill[color=gray] (1.5,-5) circle [radius=.25];	
  \draw[black,fill=white] (2.5,-7) circle [radius=.25];			
  \fill[color=gray] (3.5,-3) circle [radius=.25];	
  \draw[black,fill=white] (4.5,-7) circle [radius=.25];			
  \fill[color=gray] (5.5,-5) circle [radius=.25];	
  \fill[color=black] (6.5,-7) circle [radius=.25];	

  \fill[color=gray] (7.5,-1) circle [radius=.25];	

  \fill[color=black] (8.5,-7) circle [radius=.25];		
  \fill[color=gray] (9.5,-5) circle [radius=.25];		
  \fill[color=black] (10.5,-7) circle [radius=.25];		
  \fill[color=gray] (11.5,-3) circle [radius=.25];		
  \fill[color=black] (12.5,-7) circle [radius=.25];		
  \fill[color=gray] (13.5,-5) circle [radius=.25];		
  \fill[color=black] (14.5,-7) circle [radius=.25];		

\node at (.5,-7.8) {$\hat{u}_0$};
\node at (2.5,-7.8) {$\hat{u}_1$};
\node at (4.5,-7.8) {$\hat{u}_2$};
\node at (6.5,-7.8) {$\hat{u}_3$};
\node at (8.5,-7.8) {$\hat{u}_4$};
\node at (10.5,-7.8) {$\hat{u}_5$};
\node at (12.5,-7.8) {$\hat{u}_6$};
\node at (14.5,-7.8) {$\hat{u}_7$};

\draw [->] (8,-1.125) -- (11,-2.625) node [above=.05cm,midway,rotate=-25] {$\boldsymbol{\alpha}$};
\draw [<-] (8,-1.375) -- (11,-2.875) node [below=-.05cm,midway,rotate=-25] {$\boldsymbol{\beta}$};

\draw [->] (11.25,-3) -- (9.75,-4.5) node [above=.03cm,midway,rotate=40] {$\boldsymbol{{\alpha}^l}$};
\draw [<-] (11.25,-3.5) -- (9.75,-5) node [below=-.05cm,midway,rotate=40] {$\boldsymbol{{\beta}^l}$};

\draw [->] (11.75,-3) -- (13.25,-4.5) node [above=.03cm,midway,rotate=-40] {$\boldsymbol{{\alpha}^r}$};
\draw [<-] (11.75,-3.5) -- (13.25,-5) node [below=-.05cm,midway,rotate=-40] {$\boldsymbol{{\beta}^r}$};

\end{tikzpicture}

%% file: SCF-perf.tikz
\begin{tikzpicture}
  \pgfplotsset{
    label style = {font=\fontsize{9pt}{7.2}\selectfont},
    tick label style = {font=\fontsize{7pt}{7.2}\selectfont}
  }

\begin{axis}[
	scale = 1,
    ymode=log,
    xlabel={$E_b/N_0$ [\text{dB}]}, xlabel style={yshift=0.4em},
    ylabel={FER}, ylabel style={yshift=-0.75em},
    grid=both,
    ymajorgrids=true,
    xmajorgrids=true,
    grid style=dashed,
    width=1\columnwidth, height=6cm,
    thick,
    mark size=3,
    legend style={
      anchor={center},
      cells={anchor=west},
      column sep= 2mm,
      font=\fontsize{7pt}{7.2}\selectfont,
    },
    legend to name=SCF-perf,
    legend columns=2,
]

\addplot[
    color=red,
    mark=square,
    thick,
    mark size=3,
]
table {
1.0 7.28900e-01
1.5 3.36900e-01
2.0 8.44000e-02
2.5 1.24000e-02
3.0 1.18618e-03
3.5 1.08427e-04
4.0 1.20714e-05
};
\addlegendentry{SC}

\addplot[
    color=green,
    mark=triangle,
    thick,
    mark size=3,
]
table {
1 0.649 
1.5 0.2276 
2 0.0379
2.5 0.00322248
3 0.00014975 
3.5 7.96397e-06 
};
\addlegendentry{SC-List ($C=16$, $L=2$)}

\addplot[
    color=blue,
    mark=o,
    thick,
    mark size=3,
]
table {
1.0	6.57050e-1
1.5	2.18650e-1
2.0	2.87400e-2
2.5	1.39000e-3
3.0	6.78073e-5
3.5	3.10025e-6
};
\addlegendentry{SC-Oracle ($C=16$)}

\addplot[
    color=orange,
    mark=diamond,
    thick,
    mark size=3,
]
table {
1 0.4724 
1.5 0.1038 
2 0.0101 
2.5 0.000354035 
3 9.48661e-06 
};
\addlegendentry{SC-List ($C=16$, $L=4$)}

\addplot[
    color=black,
    mark=x,
    thick,
    mark size=3,
]
table {
1.0	7.39400e-1
1.5	3.09630e-1
2.0	5.89700e-2
2.5	4.86000e-3
3.0	2.08204e-4
3.5	6.23019e-6
};
\addlegendentry{SC-Flip ($C=16$, $T_{max}=10$)}

\addplot[
    color=magenta,
    mark=pentagon,
    thick,
    mark size=3,
]
table {
1 0.3445 
1.5 0.0504 
2 0.00311585 
2.5 6.51069e-05 
};
\addlegendentry{SC-List ($C=16$, $L=8$)}

\end{axis}
\end{tikzpicture}

%% file: errordist2.tikz
\begin{tikzpicture}
 
 \definecolor{r1}{RGB}{255,255,204}
 \definecolor{r2}{RGB}{161,218,180}
 \definecolor{r3}{RGB}{65,182,196}
 \definecolor{r4}{RGB}{44,127,184}
 \definecolor{r5}{RGB}{37,52,148}
 
  \pgfplotsset{
    label style = {font=\fontsize{9pt}{7.2}\selectfont},
    tick label style = {font=\fontsize{7pt}{7.2}\selectfont}
  }

\begin{axis}[
	width  = 9cm,
        height = 7cm,
        major x tick style = transparent,
        ybar=2*\pgflinewidth,
        bar width=7pt,
        ymajorgrids = true,
        xlabel = {Error Order},
        ylabel = {Relative frequency of occurrence},
        symbolic x coords={1,2,3+},
        xtick = data,
        scaled y ticks = false,
        enlarge x limits=0.25,
        ymin=0,
        ymax=1,
]

	\addplot[style={black,fill=r1,mark=none}]
            coordinates {(1, 0.264563) (2,0.230903) (3+,0.504534)};
            \addlegendentry{$E_b/N_0=1.0$ dB}

        \addplot[style={black,fill=r2,mark=none}]
              coordinates {(1, 0.608842) (2,0.24123) (3+,0.149928)};
             \addlegendentry{$E_b/N_0=1.5$ dB}

        \addplot[style={black,fill=r3,mark=none}]
              coordinates {(1, 0.847953) (2,0.125465) (3+,0.0265816)};
             \addlegendentry{$E_b/N_0=2.0$ dB}

        \addplot[style={black,fill=r4,mark=none}]
              coordinates {(1, 0.952793) (2,0.0430439) (3+,0.00416285)};
             \addlegendentry{$E_b/N_0=2.5$ dB}
        
        \addplot[style={black,fill=r5,mark=none}]
              coordinates {(1, 0.980672) (2,0.0185686) (3+,0.00075914)};
             \addlegendentry{$E_b/N_0=3.0$ dB}

\end{axis}
\end{tikzpicture}

%% file: CDFerrordist.tikz
\begin{tikzpicture}
  \pgfplotsset{
    label style = {font=\fontsize{9pt}{7.2}\selectfont},
    tick label style = {font=\fontsize{7pt}{7.2}\selectfont},
    yticklabel style={/pgf/number format/fixed}
  }

\begin{axis}[
	scale = 1,
    xlabel={$i$}, xlabel style={yshift=0.4em},
    ylabel={Probability of $E_1$ Occurrence}, ylabel style={yshift=-0.75em},
    grid=both,
    xmin = 0,
    xmax = 1024,
    ymin = 0,
    ymajorgrids=true,
    xmajorgrids=true,
    grid style=dashed,
    width=1\columnwidth, height=6.5cm,
    thick,
    mark size=3,
    legend style={at={(0.55,0.155)},anchor=west}
]

\addplot[
    color=green!50!black,
    dash pattern=on 2pt off 1pt,
    thick,
    mark size=3,
]
table {
0	0
1	0
2	0
3	0
4	0
5	0
6	0
7	0
8	0
9	0
10	0
11	0
12	0
13	0
14	0
15	0
16	0
17	0
18	0
19	0
20	0
21	0
22	0
23	0
24	0
25	0
26	0
27	0
28	0
29	0
30	0
31	0
32	0
33	0
34	0
35	0
36	0
37	0
38	0
39	0
40	0
41	0
42	0
43	0
44	0
45	0
46	0
47	0
48	0
49	0
50	0
51	0
52	0
53	0
54	0
55	0
56	0
57	0
58	0
59	0
60	0
61	0
62	0
63	0
64	0
65	0
66	0
67	0
68	0
69	0
70	0
71	0
72	0
73	0
74	0
75	0
76	0
77	0
78	0
79	0
80	0
81	0
82	0
83	0
84	0
85	0
86	0
87	0
88	0
89	0
90	0
91	0
92	0
93	0
94	0
95	0
96	0
97	0
98	0
99	0
100	0
101	0
102	0
103	0
104	0
105	0
106	0
107	0
108	0
109	0
110	0
111	0
112	0
113	0
114	0
115	0
116	0
117	0
118	0
119	0
120	0
121	0
122	0
123	0
124	0
125	0
126	0
127	0
128	0
129	0
130	0
131	0
132	0
133	0
134	0
135	0
136	0
137	0
138	0
139	0
140	0
141	0
142	0
143	0
144	0
145	0
146	0
147	0
148	0
149	0
150	0
151	0
152	0
153	0
154	0
155	0
156	0
157	0
158	0
159	0
160	0
161	0
162	0
163	0
164	0
165	0
166	0
167	0
168	0
169	0
170	0
171	0
172	0
173	0
174	0
175	0
176	0
177	0
178	0
179	0
180	0
181	0
182	0
183	0
184	0
185	0
186	0
187	0
188	0
189	0
190	0
191	0
192	0
193	0
194	0
195	0
196	0
197	0
198	0
199	0
200	0
201	0
202	0
203	0
204	0
205	0
206	0
207	0
208	0
209	0
210	0
211	0
212	0
213	0
214	0
215	0
216	0
217	0
218	0
219	0
220	0
221	0
222	0
223	0
224	0
225	0
226	0
227	0
228	0
229	0
230	0
231	0
232	0
233	0
234	0
235	0
236	0
237	0
238	0
239	0
240	0
241	0
242	0
243	0
244	0
245	0
246	0
247	0
248	0
249	0
250	0
251	0
252	0
253	0.10459
254	0.157365
255	0.157365
256	0.157365
257	0.157365
258	0.157365
259	0.157365
260	0.157365
261	0.157365
262	0.157365
263	0.157365
264	0.157365
265	0.157365
266	0.157365
267	0.157365
268	0.157365
269	0.157365
270	0.157365
271	0.157365
272	0.157365
273	0.157365
274	0.157365
275	0.157365
276	0.157365
277	0.157365
278	0.157365
279	0.157365
280	0.157365
281	0.157365
282	0.157365
283	0.157365
284	0.157365
285	0.157365
286	0.157365
287	0.157365
288	0.157365
289	0.157365
290	0.157365
291	0.157365
292	0.157365
293	0.157365
294	0.157365
295	0.157365
296	0.157365
297	0.157365
298	0.157365
299	0.157365
300	0.157365
301	0.157365
302	0.157365
303	0.157365
304	0.157365
305	0.157365
306	0.157365
307	0.157365
308	0.157365
309	0.157365
310	0.157365
311	0.157365
312	0.157365
313	0.157365
314	0.157365
315	0.157365
316	0.157365
317	0.157365
318	0.157365
319	0.157365
320	0.157365
321	0.157365
322	0.157365
323	0.157365
324	0.157365
325	0.157365
326	0.157365
327	0.157365
328	0.157365
329	0.157365
330	0.157365
331	0.157365
332	0.157365
333	0.157365
334	0.157365
335	0.157365
336	0.157365
337	0.157365
338	0.157365
339	0.157365
340	0.157365
341	0.157365
342	0.157365
343	0.157365
344	0.157365
345	0.157365
346	0.157365
347	0.157365
348	0.157365
349	0.157365
350	0.157365
351	0.157365
352	0.157365
353	0.157365
354	0.157365
355	0.157365
356	0.157365
357	0.157365
358	0.157365
359	0.157365
360	0.157365
361	0.157365
362	0.157365
363	0.157365
364	0.157365
365	0.157365
366	0.157365
367	0.346515
368	0.346515
369	0.346515
370	0.346515
371	0.346515
372	0.346515
373	0.346515
374	0.346515
375	0.392161
376	0.392161
377	0.392161
378	0.392161
379	0.405179
380	0.405179
381	0.409573
382	0.41142
383	0.41142
384	0.41142
385	0.41142
386	0.41142
387	0.41142
388	0.41142
389	0.41142
390	0.41142
391	0.41142
392	0.41142
393	0.41142
394	0.41142
395	0.41142
396	0.41142
397	0.41142
398	0.41142
399	0.41142
400	0.41142
401	0.41142
402	0.41142
403	0.41142
404	0.41142
405	0.41142
406	0.41142
407	0.41142
408	0.41142
409	0.41142
410	0.41142
411	0.41142
412	0.41142
413	0.41142
414	0.41142
415	0.507619
416	0.507619
417	0.507619
418	0.507619
419	0.507619
420	0.507619
421	0.507619
422	0.507619
423	0.507619
424	0.507619
425	0.507619
426	0.507619
427	0.507619
428	0.507619
429	0.507619
430	0.507619
431	0.521456
432	0.521456
433	0.521456
434	0.521456
435	0.521456
436	0.521456
437	0.521456
438	0.521456
439	0.523769
440	0.523769
441	0.523769
442	0.523769
443	0.524214
444	0.524214
445	0.524424
446	0.524447
447	0.524447
448	0.524447
449	0.524447
450	0.524447
451	0.524447
452	0.524447
453	0.524447
454	0.524447
455	0.524447
456	0.524447
457	0.524447
458	0.524447
459	0.524447
460	0.524447
461	0.524447
462	0.524447
463	0.524938
464	0.524938
465	0.524938
466	0.524938
467	0.524938
468	0.524938
469	0.524938
470	0.524938
471	0.525078
472	0.525078
473	0.525078
474	0.525078
475	0.525078
476	0.531763
477	0.531763
478	0.531763
479	0.531763
480	0.531763
481	0.531763
482	0.531763
483	0.531763
484	0.531763
485	0.546207
486	0.553592
487	0.553592
488	0.553592
489	0.558571
490	0.561118
491	0.561118
492	0.56224
493	0.56224
494	0.56224
495	0.56224
496	0.56224
497	0.563689
498	0.564437
499	0.564437
500	0.564694
501	0.564694
502	0.564694
503	0.564694
504	0.564834
505	0.564834
506	0.564834
507	0.564834
508	0.564834
509	0.564834
510	0.564834
511	0.564834
512	0.564834
513	0.564834
514	0.564834
515	0.564834
516	0.564834
517	0.564834
518	0.564834
519	0.564834
520	0.564834
521	0.564834
522	0.564834
523	0.564834
524	0.564834
525	0.564834
526	0.564834
527	0.564834
528	0.564834
529	0.564834
530	0.564834
531	0.564834
532	0.564834
533	0.564834
534	0.564834
535	0.564834
536	0.564834
537	0.564834
538	0.564834
539	0.564834
540	0.564834
541	0.564834
542	0.564834
543	0.564834
544	0.564834
545	0.564834
546	0.564834
547	0.564834
548	0.564834
549	0.564834
550	0.564834
551	0.564834
552	0.564834
553	0.564834
554	0.564834
555	0.564834
556	0.564834
557	0.564834
558	0.564834
559	0.564834
560	0.564834
561	0.564834
562	0.564834
563	0.564834
564	0.564834
565	0.564834
566	0.564834
567	0.564834
568	0.564834
569	0.564834
570	0.564834
571	0.564834
572	0.564834
573	0.564834
574	0.564834
575	0.771701
576	0.771701
577	0.771701
578	0.771701
579	0.771701
580	0.771701
581	0.771701
582	0.771701
583	0.771701
584	0.771701
585	0.771701
586	0.771701
587	0.771701
588	0.771701
589	0.771701
590	0.771701
591	0.771701
592	0.771701
593	0.771701
594	0.771701
595	0.771701
596	0.771701
597	0.771701
598	0.771701
599	0.771701
600	0.771701
601	0.771701
602	0.771701
603	0.771701
604	0.771701
605	0.771701
606	0.771701
607	0.800355
608	0.800355
609	0.800355
610	0.800355
611	0.800355
612	0.800355
613	0.800355
614	0.800355
615	0.800355
616	0.800355
617	0.800355
618	0.800355
619	0.800355
620	0.800355
621	0.800355
622	0.800355
623	0.803534
624	0.803534
625	0.803534
626	0.803534
627	0.803534
628	0.803534
629	0.803534
630	0.803534
631	0.803861
632	0.803861
633	0.803861
634	0.803861
635	0.803931
636	0.803931
637	0.803931
638	0.803931
639	0.803931
640	0.803931
641	0.803931
642	0.803931
643	0.803931
644	0.803931
645	0.803931
646	0.803931
647	0.803931
648	0.803931
649	0.803931
650	0.803931
651	0.803931
652	0.803931
653	0.803931
654	0.803931
655	0.803931
656	0.803931
657	0.803931
658	0.803931
659	0.803931
660	0.803931
661	0.803931
662	0.803931
663	0.803931
664	0.803931
665	0.803931
666	0.803931
667	0.803931
668	0.803931
669	0.803931
670	0.803931
671	0.805076
672	0.805076
673	0.805076
674	0.805076
675	0.805076
676	0.805076
677	0.805076
678	0.805076
679	0.805076
680	0.805076
681	0.805076
682	0.805076
683	0.805076
684	0.805076
685	0.805076
686	0.834034
687	0.834034
688	0.834034
689	0.834034
690	0.834034
691	0.834034
692	0.834034
693	0.848455
694	0.855116
695	0.855116
696	0.855116
697	0.859043
698	0.861193
699	0.861193
700	0.862198
701	0.862198
702	0.862198
703	0.862198
704	0.862198
705	0.862198
706	0.862198
707	0.862198
708	0.862198
709	0.862198
710	0.862198
711	0.862198
712	0.862198
713	0.862198
714	0.862198
715	0.87968
716	0.87968
717	0.886108
718	0.889567
719	0.889567
720	0.889567
721	0.889567
722	0.889567
723	0.893587
724	0.893587
725	0.894942
726	0.895667
727	0.895667
728	0.895667
729	0.896088
730	0.896181
731	0.896181
732	0.896274
733	0.896274
734	0.896274
735	0.896274
736	0.896274
737	0.896274
738	0.896274
739	0.896789
740	0.896789
741	0.896976
742	0.897022
743	0.897022
744	0.897022
745	0.897046
746	0.897093
747	0.897093
748	0.897093
749	0.897093
750	0.897093
751	0.897093
752	0.897093
753	0.897093
754	0.897093
755	0.897093
756	0.897116
757	0.897116
758	0.897116
759	0.897116
760	0.897116
761	0.897116
762	0.897116
763	0.897116
764	0.897116
765	0.897116
766	0.897116
767	0.897116
768	0.897116
769	0.897116
770	0.897116
771	0.897116
772	0.897116
773	0.897116
774	0.897116
775	0.897116
776	0.897116
777	0.897116
778	0.897116
779	0.897116
780	0.897116
781	0.897116
782	0.897116
783	0.897116
784	0.897116
785	0.897116
786	0.897116
787	0.897116
788	0.897116
789	0.897116
790	0.897116
791	0.897116
792	0.897116
793	0.897116
794	0.897116
795	0.939303
796	0.939303
797	0.954775
798	0.962371
799	0.962371
800	0.962371
801	0.962371
802	0.962371
803	0.962371
804	0.962371
805	0.962371
806	0.962371
807	0.985556
808	0.985556
809	0.985556
810	0.985556
811	0.99168
812	0.99168
813	0.993619
814	0.994531
815	0.994531
816	0.994531
817	0.994531
818	0.994531
819	0.995629
820	0.995629
821	0.996214
822	0.996447
823	0.996447
824	0.996447
825	0.996588
826	0.996658
827	0.996658
828	0.996705
829	0.996705
830	0.996705
831	0.996705
832	0.996705
833	0.996705
834	0.996705
835	0.996705
836	0.996705
837	0.996705
838	0.996705
839	0.998902
840	0.998902
841	0.998902
842	0.998902
843	0.999346
844	0.999346
845	0.999462
846	0.999579
847	0.999579
848	0.999579
849	0.999579
850	0.999579
851	0.999626
852	0.999626
853	0.99972
854	0.99972
855	0.99972
856	0.99972
857	0.99972
858	0.99972
859	0.99972
860	0.99972
861	0.99972
862	0.99972
863	0.99972
864	0.99972
865	0.99972
866	0.99972
867	0.999766
868	0.999766
869	0.999766
870	0.999766
871	0.999766
872	0.999766
873	0.999766
874	0.999766
875	0.999766
876	0.999766
877	0.999766
878	0.999766
879	0.999766
880	0.999766
881	0.999766
882	0.99979
883	0.99979
884	0.99979
885	0.99979
886	0.99979
887	0.99979
888	0.99979
889	0.99979
890	0.99979
891	0.99979
892	0.99979
893	0.99979
894	0.99979
895	0.99979
896	0.99979
897	0.99979
898	0.99979
899	0.99979
900	0.99979
901	0.99979
902	0.99979
903	0.99993
904	0.99993
905	0.99993
906	0.99993
907	0.999977
908	0.999977
909	0.999977
910	0.999977
911	0.999977
912	0.999977
913	0.999977
914	0.999977
915	0.999977
916	0.999977
917	0.999977
918	0.999977
919	0.999977
920	0.999977
921	0.999977
922	0.999977
923	0.999977
924	0.999977
925	0.999977
926	0.999977
927	0.999977
928	0.999977
929	0.999977
930	0.999977
931	0.999977
932	0.999977
933	0.999977
934	0.999977
935	0.999977
936	0.999977
937	0.999977
938	0.999977
939	0.999977
940	0.999977
941	0.999977
942	0.999977
943	0.999977
944	0.999977
945	0.999977
946	0.999977
947	0.999977
948	0.999977
949	0.999977
950	0.999977
951	0.999977
952	0.999977
953	0.999977
954	0.999977
955	0.999977
956	0.999977
957	0.999977
958	0.999977
959	0.999977
960	0.999977
961	0.999977
962	0.999977
963	0.999977
964	0.999977
965	0.999977
966	0.999977
967	0.999977
968	0.999977
969	0.999977
970	0.999977
971	0.999977
972	0.999977
973	0.999977
974	0.999977
975	0.999977
976	0.999977
977	0.999977
978	0.999977
979	0.999977
980	0.999977
981	0.999977
982	0.999977
983	0.999977
984	0.999977
985	0.999977
986	0.999977
987	0.999977
988	0.999977
989	0.999977
990	0.999977
991	0.999977
992	1
993	1
994	1
995	1
996	1
997	1
998	1
999	1
1000	1
1001	1
1002	1
1003	1
1004	1
1005	1
1006	1
1007	1
1008	1
1009	1
1010	1
1011	1
1012	1
1013	1
1014	1
1015	1
1016	1
1017	1
1018	1
1019	1
1020	1
1021	1
1022	1
1023	1
};
\addlegendentry{$PC(1024,256)$}

\addplot[
    color=blue,
    thick,
    mark size=3,
]
table {
0	0
1	0
2	0
3	0
4	0
5	0
6	0
7	0
8	0
9	0
10	0
11	0
12	0
13	0
14	0
15	0
16	0
17	0
18	0
19	0
20	0
21	0
22	0
23	0
24	0
25	0
26	0
27	0
28	0
29	0
30	0
31	0
32	0
33	0
34	0
35	0
36	0
37	0
38	0
39	0
40	0
41	0
42	0
43	0
44	0
45	0
46	0
47	0
48	0
49	0
50	0
51	0
52	0
53	0
54	0
55	0
56	0
57	0
58	0
59	0
60	0
61	0
62	0
63	0
64	0
65	0
66	0
67	0
68	0
69	0
70	0
71	0
72	0
73	0
74	0
75	0
76	0
77	0
78	0
79	0
80	0
81	0
82	0
83	0
84	0
85	0
86	0
87	0
88	0
89	0
90	0
91	0
92	0
93	0
94	0
95	0
96	0
97	0
98	0
99	0
100	0
101	0
102	0
103	0
104	0
105	0
106	0
107	0
108	0
109	0
110	0
111	0
112	0
113	0
114	0
115	0
116	0
117	0
118	0
119	0
120	0
121	0
122	0
123	0
124	0
125	0
126	0
127	0.0161897
128	0.0161897
129	0.0161897
130	0.0161897
131	0.0161897
132	0.0161897
133	0.0161897
134	0.0161897
135	0.0161897
136	0.0161897
137	0.0161897
138	0.0161897
139	0.0161897
140	0.0161897
141	0.0161897
142	0.0161897
143	0.0161897
144	0.0161897
145	0.0161897
146	0.0161897
147	0.0161897
148	0.0161897
149	0.0161897
150	0.0161897
151	0.0161897
152	0.0161897
153	0.0161897
154	0.0161897
155	0.0161897
156	0.0161897
157	0.0161897
158	0.0161897
159	0.0161897
160	0.0161897
161	0.0161897
162	0.0161897
163	0.0161897
164	0.0161897
165	0.0161897
166	0.0161897
167	0.0161897
168	0.0161897
169	0.0161897
170	0.0161897
171	0.0161897
172	0.0161897
173	0.0161897
174	0.0161897
175	0.0161897
176	0.0161897
177	0.0161897
178	0.0161897
179	0.0161897
180	0.0161897
181	0.0161897
182	0.0161897
183	0.0161897
184	0.0161897
185	0.0161897
186	0.0161897
187	0.0161897
188	0.0161897
189	0.0161897
190	0.0161897
191	0.0166993
192	0.0166993
193	0.0166993
194	0.0166993
195	0.0166993
196	0.0166993
197	0.0166993
198	0.0166993
199	0.0166993
200	0.0166993
201	0.0166993
202	0.0166993
203	0.0166993
204	0.0166993
205	0.0166993
206	0.0166993
207	0.0166993
208	0.0166993
209	0.0166993
210	0.0166993
211	0.0166993
212	0.0166993
213	0.0166993
214	0.0166993
215	0.0166993
216	0.0166993
217	0.0166993
218	0.0166993
219	0.0166993
220	0.0166993
221	0.0413563
222	0.052646
223	0.052646
224	0.052646
225	0.052646
226	0.052646
227	0.052646
228	0.052646
229	0.052646
230	0.052646
231	0.0894159
232	0.0894159
233	0.0894159
234	0.0894159
235	0.098628
236	0.098628
237	0.101529
238	0.102862
239	0.102862
240	0.102862
241	0.102862
242	0.102862
243	0.104547
244	0.104547
245	0.105135
246	0.105449
247	0.105449
248	0.105449
249	0.105723
250	0.105919
251	0.105919
252	0.105958
253	0.105958
254	0.105958
255	0.105958
256	0.105958
257	0.105958
258	0.105958
259	0.105958
260	0.105958
261	0.105958
262	0.105958
263	0.105958
264	0.105958
265	0.105958
266	0.105958
267	0.105958
268	0.105958
269	0.105958
270	0.105958
271	0.105958
272	0.105958
273	0.105958
274	0.105958
275	0.105958
276	0.105958
277	0.105958
278	0.105958
279	0.105958
280	0.105958
281	0.105958
282	0.105958
283	0.105958
284	0.105958
285	0.105958
286	0.105958
287	0.105958
288	0.105958
289	0.105958
290	0.105958
291	0.105958
292	0.105958
293	0.105958
294	0.105958
295	0.105958
296	0.105958
297	0.105958
298	0.105958
299	0.105958
300	0.105958
301	0.105958
302	0.105958
303	0.105958
304	0.105958
305	0.105958
306	0.105958
307	0.105958
308	0.105958
309	0.105958
310	0.105958
311	0.105958
312	0.105958
313	0.105958
314	0.105958
315	0.128695
316	0.128695
317	0.13775
318	0.141631
319	0.141631
320	0.141631
321	0.141631
322	0.141631
323	0.141631
324	0.141631
325	0.141631
326	0.141631
327	0.141631
328	0.141631
329	0.141631
330	0.141631
331	0.141631
332	0.141631
333	0.141631
334	0.141631
335	0.181693
336	0.181693
337	0.181693
338	0.181693
339	0.181693
340	0.181693
341	0.181693
342	0.181693
343	0.188044
344	0.188044
345	0.188044
346	0.188044
347	0.18969
348	0.18969
349	0.190043
350	0.190435
351	0.190435
352	0.190435
353	0.190435
354	0.190435
355	0.190435
356	0.190435
357	0.190435
358	0.190435
359	0.191258
360	0.191258
361	0.191258
362	0.191258
363	0.191298
364	0.218189
365	0.218189
366	0.218189
367	0.218189
368	0.218189
369	0.251509
370	0.26813
371	0.26813
372	0.275931
373	0.275931
374	0.275931
375	0.275931
376	0.279929
377	0.279929
378	0.279929
379	0.279929
380	0.279929
381	0.279929
382	0.279929
383	0.279929
384	0.279929
385	0.279929
386	0.279929
387	0.279929
388	0.279929
389	0.279929
390	0.279929
391	0.279929
392	0.279929
393	0.279929
394	0.279929
395	0.279929
396	0.279929
397	0.279929
398	0.279929
399	0.281458
400	0.281458
401	0.281458
402	0.281458
403	0.281458
404	0.281458
405	0.281458
406	0.281458
407	0.281458
408	0.281458
409	0.324931
410	0.345708
411	0.345708
412	0.356409
413	0.356409
414	0.356409
415	0.356409
416	0.356409
417	0.356409
418	0.356409
419	0.356409
420	0.356409
421	0.381693
422	0.39459
423	0.39459
424	0.39459
425	0.403567
426	0.408232
427	0.408232
428	0.410192
429	0.410192
430	0.410192
431	0.410192
432	0.410192
433	0.41274
434	0.4147
435	0.4147
436	0.415406
437	0.415406
438	0.415406
439	0.415406
440	0.415719
441	0.415719
442	0.415719
443	0.415719
444	0.415719
445	0.415719
446	0.415719
447	0.415719
448	0.415719
449	0.415719
450	0.415719
451	0.426382
452	0.426382
453	0.430968
454	0.433007
455	0.433007
456	0.433007
457	0.434339
458	0.434888
459	0.434888
460	0.435241
461	0.435241
462	0.435241
463	0.435241
464	0.435241
465	0.435633
466	0.435711
467	0.435711
468	0.435868
469	0.435868
470	0.435868
471	0.435868
472	0.435907
473	0.435907
474	0.435907
475	0.435907
476	0.435907
477	0.435907
478	0.435907
479	0.435907
480	0.435907
481	0.435986
482	0.435986
483	0.435986
484	0.435986
485	0.435986
486	0.435986
487	0.435986
488	0.435986
489	0.435986
490	0.435986
491	0.435986
492	0.435986
493	0.435986
494	0.435986
495	0.435986
496	0.436025
497	0.436025
498	0.436025
499	0.436025
500	0.436025
501	0.436025
502	0.436025
503	0.436025
504	0.436025
505	0.436025
506	0.436025
507	0.436025
508	0.436025
509	0.436025
510	0.436025
511	0.436025
512	0.436025
513	0.436025
514	0.436025
515	0.436025
516	0.436025
517	0.436025
518	0.436025
519	0.436025
520	0.436025
521	0.436025
522	0.436025
523	0.436025
524	0.436025
525	0.436025
526	0.436025
527	0.436025
528	0.436025
529	0.436025
530	0.436025
531	0.436025
532	0.436025
533	0.436025
534	0.436025
535	0.436025
536	0.436025
537	0.436025
538	0.436025
539	0.436025
540	0.436025
541	0.436025
542	0.436025
543	0.487887
544	0.487887
545	0.487887
546	0.487887
547	0.487887
548	0.487887
549	0.487887
550	0.487887
551	0.487887
552	0.487887
553	0.487887
554	0.487887
555	0.487887
556	0.487887
557	0.487887
558	0.487887
559	0.493375
560	0.493375
561	0.493375
562	0.493375
563	0.493375
564	0.493375
565	0.493375
566	0.493375
567	0.493963
568	0.493963
569	0.493963
570	0.493963
571	0.494159
572	0.523246
573	0.523246
574	0.523285
575	0.523285
576	0.523285
577	0.523285
578	0.523285
579	0.523285
580	0.523285
581	0.523285
582	0.523285
583	0.523285
584	0.523285
585	0.523285
586	0.523285
587	0.523285
588	0.523285
589	0.523285
590	0.523285
591	0.523559
592	0.523559
593	0.523559
594	0.523559
595	0.523559
596	0.523559
597	0.564955
598	0.587103
599	0.587103
600	0.587103
601	0.601764
602	0.609016
603	0.609016
604	0.61223
605	0.61223
606	0.61223
607	0.61223
608	0.61223
609	0.61223
610	0.61223
611	0.634496
612	0.634496
613	0.642023
614	0.645786
615	0.645786
616	0.645786
617	0.648216
618	0.649745
619	0.649745
620	0.650608
621	0.650608
622	0.650608
623	0.650608
624	0.650608
625	0.651392
626	0.651823
627	0.651823
628	0.652019
629	0.652019
630	0.652019
631	0.652019
632	0.652136
633	0.652136
634	0.652136
635	0.652136
636	0.652136
637	0.652136
638	0.652136
639	0.652136
640	0.652136
641	0.652136
642	0.652136
643	0.652136
644	0.652136
645	0.652136
646	0.652136
647	0.652136
648	0.652136
649	0.652136
650	0.652136
651	0.692082
652	0.692082
653	0.70737
654	0.715367
655	0.715367
656	0.715367
657	0.715367
658	0.715367
659	0.725559
660	0.725559
661	0.728303
662	0.730027
663	0.730027
664	0.730027
665	0.731243
666	0.73187
667	0.73187
668	0.731948
669	0.731948
670	0.731948
671	0.731948
672	0.731948
673	0.731948
674	0.731948
675	0.733673
676	0.733673
677	0.734143
678	0.734418
679	0.734418
680	0.734418
681	0.734614
682	0.73481
683	0.73481
684	0.734849
685	0.734849
686	0.734849
687	0.734849
688	0.766092
689	0.766092
690	0.766092
691	0.766092
692	0.766092
693	0.766092
694	0.766092
695	0.766092
696	0.766092
697	0.766092
698	0.766092
699	0.766092
700	0.766092
701	0.766092
702	0.766092
703	0.766092
704	0.766092
705	0.766092
706	0.766092
707	0.76617
708	0.801921
709	0.80196
710	0.80196
711	0.80196
712	0.820815
713	0.820815
714	0.820815
715	0.820815
716	0.820815
717	0.820815
718	0.820815
719	0.820815
720	0.829518
721	0.829518
722	0.829518
723	0.829518
724	0.829518
725	0.829518
726	0.829518
727	0.829518
728	0.829518
729	0.829518
730	0.829518
731	0.829518
732	0.829518
733	0.829518
734	0.829518
735	0.829518
736	0.834418
737	0.834418
738	0.834418
739	0.834418
740	0.834418
741	0.834418
742	0.834418
743	0.834418
744	0.834418
745	0.834418
746	0.834418
747	0.834418
748	0.834418
749	0.834418
750	0.834418
751	0.834418
752	0.834418
753	0.834418
754	0.834418
755	0.834418
756	0.834418
757	0.834418
758	0.834418
759	0.834418
760	0.834418
761	0.834418
762	0.834418
763	0.834418
764	0.834418
765	0.834418
766	0.834418
767	0.834418
768	0.834418
769	0.834418
770	0.834418
771	0.834418
772	0.834418
773	0.834418
774	0.834418
775	0.845472
776	0.845472
777	0.845472
778	0.845472
779	0.847628
780	0.847628
781	0.848608
782	0.848844
783	0.848844
784	0.848844
785	0.848844
786	0.848844
787	0.849118
788	0.849118
789	0.849236
790	0.849275
791	0.849275
792	0.878401
793	0.878401
794	0.878401
795	0.878401
796	0.878401
797	0.878401
798	0.878401
799	0.878401
800	0.878401
801	0.878401
802	0.913995
803	0.913995
804	0.932536
805	0.932536
806	0.932536
807	0.932536
808	0.941592
809	0.941592
810	0.941592
811	0.941592
812	0.941592
813	0.941592
814	0.941592
815	0.941592
816	0.945982
817	0.945982
818	0.945982
819	0.945982
820	0.945982
821	0.945982
822	0.945982
823	0.945982
824	0.945982
825	0.945982
826	0.945982
827	0.945982
828	0.945982
829	0.945982
830	0.945982
831	0.945982
832	0.945982
833	0.968091
834	0.979263
835	0.979263
836	0.983457
837	0.983457
838	0.983457
839	0.983457
840	0.98577
841	0.98577
842	0.98577
843	0.98577
844	0.98577
845	0.98577
846	0.98577
847	0.98577
848	0.987221
849	0.987221
850	0.987221
851	0.987221
852	0.987221
853	0.987221
854	0.987221
855	0.987221
856	0.987221
857	0.987221
858	0.987221
859	0.987221
860	0.987221
861	0.987221
862	0.987221
863	0.987221
864	0.98777
865	0.98777
866	0.98777
867	0.98777
868	0.98777
869	0.98777
870	0.98777
871	0.98777
872	0.98777
873	0.98777
874	0.98777
875	0.98777
876	0.98777
877	0.98777
878	0.98777
879	0.98777
880	0.98777
881	0.98777
882	0.98777
883	0.98777
884	0.98777
885	0.98777
886	0.98777
887	0.98777
888	0.98777
889	0.98777
890	0.98777
891	0.98777
892	0.98777
893	0.98777
894	0.98777
895	0.98777
896	0.98777
897	0.994355
898	0.997217
899	0.997217
900	0.998746
901	0.998746
902	0.998746
903	0.998746
904	0.999569
905	0.999569
906	0.999569
907	0.999569
908	0.999569
909	0.999569
910	0.999569
911	0.999569
912	0.999922
913	0.999922
914	0.999922
915	0.999922
916	0.999922
917	0.999922
918	0.999922
919	0.999922
920	0.999922
921	0.999922
922	0.999922
923	0.999922
924	0.999922
925	0.999922
926	0.999922
927	0.999922
928	1
929	1
930	1
931	1
932	1
933	1
934	1
935	1
936	1
937	1
938	1
939	1
940	1
941	1
942	1
943	1
944	1
945	1
946	1
947	1
948	1
949	1
950	1
951	1
952	1
953	1
954	1
955	1
956	1
957	1
958	1
959	1
960	1
961	1
962	1
963	1
964	1
965	1
966	1
967	1
968	1
969	1
970	1
971	1
972	1
973	1
974	1
975	1
976	1
977	1
978	1
979	1
980	1
981	1
982	1
983	1
984	1
985	1
986	1
987	1
988	1
989	1
990	1
991	1
992	1
993	1
994	1
995	1
996	1
997	1
998	1
999	1
1000	1
1001	1
1002	1
1003	1
1004	1
1005	1
1006	1
1007	1
1008	1
1009	1
1010	1
1011	1
1012	1
1013	1
1014	1
1015	1
1016	1
1017	1
1018	1
1019	1
1020	1
1021	1
1022	1
1023	1
};
\addlegendentry{$PC(1024,512)$}

\addplot[
    color=red,
    dashed,
    thick,
    mark size=3,
]
table {
0	0
1	0
2	0
3	0
4	0
5	0
6	0
7	0
8	0
9	0
10	0
11	0
12	0
13	0
14	0
15	0
16	0
17	0
18	0
19	0
20	0
21	0
22	0
23	0
24	0
25	0
26	0
27	0
28	0
29	0
30	0
31	0
32	0
33	0
34	0
35	0
36	0
37	0
38	0
39	0
40	0
41	0
42	0
43	0
44	0
45	0
46	0
47	0
48	0
49	0
50	0
51	0
52	0
53	0
54	0
55	0
56	0
57	0
58	0
59	0
60	0
61	0
62	0
63	0.0261106
64	0.0261106
65	0.0261106
66	0.0261106
67	0.0261106
68	0.0261106
69	0.0261106
70	0.0261106
71	0.0261106
72	0.0261106
73	0.0261106
74	0.0261106
75	0.0261106
76	0.0261106
77	0.0261106
78	0.0261106
79	0.0261106
80	0.0261106
81	0.0261106
82	0.0261106
83	0.0261106
84	0.0261106
85	0.0261106
86	0.0261106
87	0.0261106
88	0.0261106
89	0.0261106
90	0.0261106
91	0.0261106
92	0.0261106
93	0.0261106
94	0.108123
95	0.109222
96	0.109222
97	0.109222
98	0.109222
99	0.109222
100	0.109222
101	0.109222
102	0.109222
103	0.109222
104	0.109222
105	0.109222
106	0.109222
107	0.109222
108	0.109222
109	0.151963
110	0.173233
111	0.173255
112	0.173255
113	0.173255
114	0.173255
115	0.20311
116	0.20311
117	0.215287
118	0.221141
119	0.221141
120	0.221141
121	0.225822
122	0.227947
123	0.227948
124	0.228976
125	0.228976
126	0.228976
127	0.228976
128	0.228976
129	0.228976
130	0.228976
131	0.228976
132	0.228976
133	0.228976
134	0.228976
135	0.228976
136	0.228976
137	0.228976
138	0.228976
139	0.228976
140	0.228976
141	0.228976
142	0.228976
143	0.228976
144	0.228976
145	0.228976
146	0.228976
147	0.228976
148	0.228976
149	0.228976
150	0.228976
151	0.228976
152	0.228976
153	0.228976
154	0.228976
155	0.27386
156	0.27386
157	0.293542
158	0.303053
159	0.303053
160	0.303053
161	0.303053
162	0.303053
163	0.303053
164	0.303053
165	0.303053
166	0.303053
167	0.3316
168	0.3316
169	0.3316
170	0.3316
171	0.341851
172	0.341851
173	0.345802
174	0.34777
175	0.34777
176	0.34777
177	0.34777
178	0.34777
179	0.350532
180	0.385578
181	0.385758
182	0.385874
183	0.385874
184	0.403208
185	0.40326
186	0.40328
187	0.40328
188	0.403285
189	0.403285
190	0.403285
191	0.403285
192	0.403285
193	0.403285
194	0.403285
195	0.403285
196	0.403285
197	0.403285
198	0.403285
199	0.408152
200	0.408152
201	0.408152
202	0.408152
203	0.409523
204	0.432368
205	0.432439
206	0.432486
207	0.432486
208	0.432486
209	0.460162
210	0.473628
211	0.473634
212	0.479845
213	0.479847
214	0.479847
215	0.479847
216	0.482791
217	0.482791
218	0.482791
219	0.482791
220	0.482791
221	0.482791
222	0.482791
223	0.482791
224	0.482791
225	0.492973
226	0.497923
227	0.497925
228	0.500126
229	0.500126
230	0.500126
231	0.500126
232	0.50126
233	0.50126
234	0.50126
235	0.50126
236	0.50126
237	0.50126
238	0.50126
239	0.50126
240	0.501686
241	0.501686
242	0.501686
243	0.501686
244	0.501686
245	0.501686
246	0.501686
247	0.501686
248	0.501686
249	0.501686
250	0.501686
251	0.501686
252	0.501686
253	0.501686
254	0.501686
255	0.501686
256	0.501686
257	0.501686
258	0.501686
259	0.501686
260	0.501686
261	0.501686
262	0.501686
263	0.501686
264	0.501686
265	0.501686
266	0.501686
267	0.501686
268	0.501686
269	0.501686
270	0.501686
271	0.528992
272	0.528992
273	0.528992
274	0.528992
275	0.528992
276	0.528992
277	0.528992
278	0.528992
279	0.537462
280	0.537462
281	0.537462
282	0.537462
283	0.540241
284	0.540241
285	0.541283
286	0.541762
287	0.541762
288	0.541762
289	0.541762
290	0.541762
291	0.541762
292	0.541762
293	0.541762
294	0.541762
295	0.543497
296	0.543497
297	0.576326
298	0.592811
299	0.592816
300	0.600861
301	0.600861
302	0.600861
303	0.600861
304	0.600861
305	0.613936
306	0.620145
307	0.620148
308	0.623324
309	0.623324
310	0.623324
311	0.623324
312	0.624523
313	0.624523
314	0.624523
315	0.624523
316	0.624523
317	0.624523
318	0.624523
319	0.624523
320	0.624523
321	0.624523
322	0.624523
323	0.657707
324	0.657707
325	0.67264
326	0.680175
327	0.680175
328	0.680175
329	0.686188
330	0.689162
331	0.689162
332	0.690668
333	0.690668
334	0.690668
335	0.690668
336	0.690668
337	0.69288
338	0.694056
339	0.694056
340	0.694644
341	0.694644
342	0.694644
343	0.694644
344	0.694921
345	0.694921
346	0.694921
347	0.694921
348	0.694921
349	0.694921
350	0.694921
351	0.694921
352	0.71732
353	0.717425
354	0.717472
355	0.717472
356	0.717505
357	0.717505
358	0.717505
359	0.717505
360	0.717519
361	0.717519
362	0.717519
363	0.717519
364	0.717519
365	0.717519
366	0.717519
367	0.717519
368	0.717526
369	0.717526
370	0.717526
371	0.717526
372	0.717526
373	0.717526
374	0.717526
375	0.717526
376	0.717526
377	0.717526
378	0.717526
379	0.717526
380	0.717526
381	0.717526
382	0.717526
383	0.717526
384	0.717526
385	0.717526
386	0.717526
387	0.724998
388	0.724998
389	0.728153
390	0.729716
391	0.729716
392	0.759026
393	0.759201
394	0.759293
395	0.759293
396	0.759358
397	0.759358
398	0.759358
399	0.759358
400	0.773993
401	0.774049
402	0.774076
403	0.774076
404	0.774081
405	0.774081
406	0.774081
407	0.774081
408	0.774081
409	0.774081
410	0.774081
411	0.774081
412	0.774081
413	0.774081
414	0.774081
415	0.774081
416	0.780716
417	0.780728
418	0.780734
419	0.780734
420	0.780734
421	0.780734
422	0.780734
423	0.780734
424	0.780734
425	0.780734
426	0.780734
427	0.780734
428	0.780734
429	0.780734
430	0.780734
431	0.780734
432	0.780734
433	0.780734
434	0.780734
435	0.780734
436	0.780734
437	0.780734
438	0.780734
439	0.780734
440	0.780734
441	0.780734
442	0.780734
443	0.780734
444	0.780734
445	0.780734
446	0.780734
447	0.780734
448	0.783973
449	0.783976
450	0.783977
451	0.783977
452	0.783977
453	0.783977
454	0.783977
455	0.783977
456	0.783977
457	0.783977
458	0.783977
459	0.783977
460	0.783977
461	0.783977
462	0.783977
463	0.783977
464	0.783977
465	0.783977
466	0.783977
467	0.783977
468	0.783977
469	0.783977
470	0.783977
471	0.783977
472	0.783977
473	0.783977
474	0.783977
475	0.783977
476	0.783977
477	0.783977
478	0.783977
479	0.783977
480	0.783977
481	0.783977
482	0.783977
483	0.783977
484	0.783977
485	0.783977
486	0.783977
487	0.783977
488	0.783977
489	0.783977
490	0.783977
491	0.783977
492	0.783977
493	0.783977
494	0.783977
495	0.783977
496	0.783977
497	0.783977
498	0.783977
499	0.783977
500	0.783977
501	0.783977
502	0.783977
503	0.783977
504	0.783977
505	0.783977
506	0.783977
507	0.783977
508	0.783977
509	0.783977
510	0.783977
511	0.783977
512	0.783977
513	0.783977
514	0.783977
515	0.783977
516	0.783977
517	0.783977
518	0.783977
519	0.783977
520	0.783977
521	0.783977
522	0.783977
523	0.783977
524	0.783977
525	0.783977
526	0.813475
527	0.813746
528	0.813746
529	0.813746
530	0.813746
531	0.813746
532	0.813746
533	0.832975
534	0.841855
535	0.841858
536	0.841858
537	0.849413
538	0.852904
539	0.852904
540	0.854432
541	0.854432
542	0.854432
543	0.854432
544	0.854432
545	0.854432
546	0.854432
547	0.86563
548	0.86563
549	0.870131
550	0.872285
551	0.872285
552	0.872285
553	0.874058
554	0.874869
555	0.874869
556	0.875301
557	0.875301
558	0.875301
559	0.875301
560	0.89419
561	0.894263
562	0.894313
563	0.894313
564	0.894334
565	0.894334
566	0.894334
567	0.894334
568	0.89434
569	0.89434
570	0.89434
571	0.89434
572	0.89434
573	0.89434
574	0.89434
575	0.89434
576	0.89434
577	0.89434
578	0.89434
579	0.896728
580	0.920895
581	0.921086
582	0.921167
583	0.921167
584	0.93314
585	0.933182
586	0.933215
587	0.933215
588	0.933218
589	0.933218
590	0.933218
591	0.933218
592	0.938876
593	0.938891
594	0.938892
595	0.938892
596	0.938892
597	0.938892
598	0.938892
599	0.938892
600	0.938892
601	0.938892
602	0.938892
603	0.938892
604	0.938892
605	0.938892
606	0.938892
607	0.938892
608	0.941761
609	0.941761
610	0.941761
611	0.941761
612	0.941761
613	0.941761
614	0.941761
615	0.941761
616	0.941761
617	0.941761
618	0.941761
619	0.941761
620	0.941761
621	0.941761
622	0.941761
623	0.941761
624	0.941761
625	0.941761
626	0.941761
627	0.941761
628	0.941761
629	0.941761
630	0.941761
631	0.941761
632	0.941761
633	0.941761
634	0.941761
635	0.941761
636	0.941761
637	0.941761
638	0.941761
639	0.941761
640	0.941761
641	0.964307
642	0.975562
643	0.975573
644	0.980789
645	0.980789
646	0.980792
647	0.980792
648	0.983394
649	0.983394
650	0.983394
651	0.983394
652	0.983394
653	0.983394
654	0.983394
655	0.983394
656	0.984676
657	0.984676
658	0.984676
659	0.984676
660	0.984676
661	0.984676
662	0.984676
663	0.984676
664	0.984676
665	0.984676
666	0.984676
667	0.984676
668	0.984676
669	0.984676
670	0.984676
671	0.984676
672	0.985287
673	0.985287
674	0.985287
675	0.985287
676	0.985287
677	0.985287
678	0.985287
679	0.985287
680	0.985287
681	0.985287
682	0.985287
683	0.985287
684	0.985287
685	0.985287
686	0.985287
687	0.985287
688	0.985287
689	0.985287
690	0.985287
691	0.985287
692	0.985287
693	0.985287
694	0.985287
695	0.985287
696	0.985287
697	0.985287
698	0.985287
699	0.985287
700	0.985287
701	0.985287
702	0.985287
703	0.985287
704	0.985564
705	0.985564
706	0.985564
707	0.985564
708	0.985564
709	0.985564
710	0.985564
711	0.985564
712	0.985564
713	0.985564
714	0.985564
715	0.985564
716	0.985564
717	0.985564
718	0.985564
719	0.985564
720	0.985564
721	0.985564
722	0.985564
723	0.985564
724	0.985564
725	0.985564
726	0.985564
727	0.985564
728	0.985564
729	0.985564
730	0.985564
731	0.985564
732	0.985564
733	0.985564
734	0.985564
735	0.985564
736	0.985564
737	0.985564
738	0.985564
739	0.985564
740	0.985564
741	0.985564
742	0.985564
743	0.985564
744	0.985564
745	0.985564
746	0.985564
747	0.985564
748	0.985564
749	0.985564
750	0.985564
751	0.985564
752	0.985564
753	0.985564
754	0.985564
755	0.985564
756	0.985564
757	0.985564
758	0.985564
759	0.985564
760	0.985564
761	0.985564
762	0.985564
763	0.985564
764	0.985564
765	0.985564
766	0.985564
767	0.985564
768	0.985564
769	0.992927
770	0.996621
771	0.996622
772	0.99841
773	0.99841
774	0.99841
775	0.99841
776	0.999318
777	0.999318
778	0.999318
779	0.999318
780	0.999318
781	0.999318
782	0.999318
783	0.999318
784	0.999707
785	0.999707
786	0.999707
787	0.999707
788	0.999707
789	0.999707
790	0.999707
791	0.999707
792	0.999707
793	0.999707
794	0.999707
795	0.999707
796	0.999707
797	0.999707
798	0.999707
799	0.999707
800	0.999891
801	0.999891
802	0.999891
803	0.999891
804	0.999891
805	0.999891
806	0.999891
807	0.999891
808	0.999891
809	0.999891
810	0.999891
811	0.999891
812	0.999891
813	0.999891
814	0.999891
815	0.999891
816	0.999891
817	0.999891
818	0.999891
819	0.999891
820	0.999891
821	0.999891
822	0.999891
823	0.999891
824	0.999891
825	0.999891
826	0.999891
827	0.999891
828	0.999891
829	0.999891
830	0.999891
831	0.999891
832	0.999956
833	0.999956
834	0.999956
835	0.999956
836	0.999956
837	0.999956
838	0.999956
839	0.999956
840	0.999956
841	0.999956
842	0.999956
843	0.999956
844	0.999956
845	0.999956
846	0.999956
847	0.999956
848	0.999956
849	0.999956
850	0.999956
851	0.999956
852	0.999956
853	0.999956
854	0.999956
855	0.999956
856	0.999956
857	0.999956
858	0.999956
859	0.999956
860	0.999956
861	0.999956
862	0.999956
863	0.999956
864	0.999956
865	0.999956
866	0.999956
867	0.999956
868	0.999956
869	0.999956
870	0.999956
871	0.999956
872	0.999956
873	0.999956
874	0.999956
875	0.999956
876	0.999956
877	0.999956
878	0.999956
879	0.999956
880	0.999956
881	0.999956
882	0.999956
883	0.999956
884	0.999956
885	0.999956
886	0.999956
887	0.999956
888	0.999956
889	0.999956
890	0.999956
891	0.999956
892	0.999956
893	0.999956
894	0.999956
895	0.999956
896	1
897	1
898	1
899	1
900	1
901	1
902	1
903	1
904	1
905	1
906	1
907	1
908	1
909	1
910	1
911	1
912	1
913	1
914	1
915	1
916	1
917	1
918	1
919	1
920	1
921	1
922	1
923	1
924	1
925	1
926	1
927	1
928	1
929	1
930	1
931	1
932	1
933	1
934	1
935	1
936	1
937	1
938	1
939	1
940	1
941	1
942	1
943	1
944	1
945	1
946	1
947	1
948	1
949	1
950	1
951	1
952	1
953	1
954	1
955	1
956	1
957	1
958	1
959	1
960	1
961	1
962	1
963	1
964	1
965	1
966	1
967	1
968	1
969	1
970	1
971	1
972	1
973	1
974	1
975	1
976	1
977	1
978	1
979	1
980	1
981	1
982	1
983	1
984	1
985	1
986	1
987	1
988	1
989	1
990	1
991	1
992	1
993	1
994	1
995	1
996	1
997	1
998	1
999	1
1000	1
1001	1
1002	1
1003	1
1004	1
1005	1
1006	1
1007	1
1008	1
1009	1
1010	1
1011	1
1012	1
1013	1
1014	1
1015	1
1016	1
1017	1
1018	1
1019	1
1020	1
1021	1
1022	1
1023	1
};
\addlegendentry{$PC(1024,768)$}
\end{axis}
\end{tikzpicture}

%% file: CDFerrordist-ebn0.tikz
\begin{tikzpicture}
  \pgfplotsset{
    label style = {font=\fontsize{9pt}{7.2}\selectfont},
    tick label style = {font=\fontsize{7pt}{7.2}\selectfont},
    yticklabel style={/pgf/number format/fixed}
  }

\begin{axis}[
	scale = 1,
    xlabel={$i$}, xlabel style={yshift=0.4em},
    ylabel={Probability of $E_1$ Occurrence}, ylabel style={yshift=-0.75em},
    grid=both,
    xmin = 0,
    xmax = 1024,
    ymin = 0,
    ymajorgrids=true,
    xmajorgrids=true,
    grid style=dashed,
    width=1\columnwidth, height=6.5cm,
    thick,
    mark size=3,
    legend style={at={(0.01,0.8)},anchor=west}
]

\addplot[
    color=green!50!black,
    dash pattern=on 2pt off 1pt,
    thick,
    mark size=3,
]
table {
0	0
1	0
2	0
3	0
4	0
5	0
6	0
7	0
8	0
9	0
10	0
11	0
12	0
13	0
14	0
15	0
16	0
17	0
18	0
19	0
20	0
21	0
22	0
23	0
24	0
25	0
26	0
27	0
28	0
29	0
30	0
31	0
32	0
33	0
34	0
35	0
36	0
37	0
38	0
39	0
40	0
41	0
42	0
43	0
44	0
45	0
46	0
47	0
48	0
49	0
50	0
51	0
52	0
53	0
54	0
55	0
56	0
57	0
58	0
59	0
60	0
61	0
62	0
63	0
64	0
65	0
66	0
67	0
68	0
69	0
70	0
71	0
72	0
73	0
74	0
75	0
76	0
77	0
78	0
79	0
80	0
81	0
82	0
83	0
84	0
85	0
86	0
87	0
88	0
89	0
90	0
91	0
92	0
93	0
94	0
95	0
96	0
97	0
98	0
99	0
100	0
101	0
102	0
103	0
104	0
105	0
106	0
107	0
108	0
109	0
110	0
111	0
112	0
113	0
114	0
115	0
116	0
117	0
118	0
119	0
120	0
121	0
122	0
123	0
124	0
125	0
126	0
127	0.0382357
128	0.0382357
129	0.0382357
130	0.0382357
131	0.0382357
132	0.0382357
133	0.0382357
134	0.0382357
135	0.0382357
136	0.0382357
137	0.0382357
138	0.0382357
139	0.0382357
140	0.0382357
141	0.0382357
142	0.0382357
143	0.0382357
144	0.0382357
145	0.0382357
146	0.0382357
147	0.0382357
148	0.0382357
149	0.0382357
150	0.0382357
151	0.0382357
152	0.0382357
153	0.0382357
154	0.0382357
155	0.0382357
156	0.0382357
157	0.0382357
158	0.0382357
159	0.0382357
160	0.0382357
161	0.0382357
162	0.0382357
163	0.0382357
164	0.0382357
165	0.0382357
166	0.0382357
167	0.0382357
168	0.0382357
169	0.0382357
170	0.0382357
171	0.0382357
172	0.0382357
173	0.0382357
174	0.0382357
175	0.0382357
176	0.0382357
177	0.0382357
178	0.0382357
179	0.0382357
180	0.0382357
181	0.0382357
182	0.0382357
183	0.0382357
184	0.0382357
185	0.0382357
186	0.0382357
187	0.0382357
188	0.0382357
189	0.0382357
190	0.0382357
191	0.0442998
192	0.0442998
193	0.0442998
194	0.0442998
195	0.0442998
196	0.0442998
197	0.0442998
198	0.0442998
199	0.0442998
200	0.0442998
201	0.0442998
202	0.0442998
203	0.0442998
204	0.0442998
205	0.0442998
206	0.0442998
207	0.0442998
208	0.0442998
209	0.0442998
210	0.0442998
211	0.0442998
212	0.0442998
213	0.0442998
214	0.0442998
215	0.0442998
216	0.0442998
217	0.0442998
218	0.0442998
219	0.0442998
220	0.0442998
221	0.0939909
222	0.118935
223	0.118957
224	0.118957
225	0.118957
226	0.118957
227	0.118957
228	0.118957
229	0.118957
230	0.118957
231	0.179202
232	0.179202
233	0.179202
234	0.179202
235	0.202654
236	0.202654
237	0.212179
238	0.217
239	0.217
240	0.217
241	0.217
242	0.217
243	0.223968
244	0.223968
245	0.226563
246	0.227858
247	0.227858
248	0.227858
249	0.228801
250	0.229273
251	0.229273
252	0.229492
253	0.229492
254	0.229492
255	0.229492
256	0.229492
257	0.229492
258	0.229492
259	0.229492
260	0.229492
261	0.229492
262	0.229492
263	0.229492
264	0.229492
265	0.229492
266	0.229492
267	0.229492
268	0.229492
269	0.229492
270	0.229492
271	0.229492
272	0.229492
273	0.229492
274	0.229492
275	0.229492
276	0.229492
277	0.229492
278	0.229492
279	0.229492
280	0.229492
281	0.229492
282	0.229492
283	0.229492
284	0.229492
285	0.229492
286	0.229492
287	0.229492
288	0.229492
289	0.229492
290	0.229492
291	0.229492
292	0.229492
293	0.229492
294	0.229492
295	0.229492
296	0.229492
297	0.229492
298	0.229492
299	0.229492
300	0.229492
301	0.229492
302	0.229492
303	0.229492
304	0.229492
305	0.229492
306	0.229492
307	0.229492
308	0.229492
309	0.229492
310	0.229492
311	0.229492
312	0.229492
313	0.229492
314	0.229492
315	0.27497
316	0.27497
317	0.294978
318	0.304998
319	0.304998
320	0.304998
321	0.304998
322	0.304998
323	0.304998
324	0.304998
325	0.304998
326	0.304998
327	0.304998
328	0.304998
329	0.304998
330	0.304998
331	0.304998
332	0.304998
333	0.304998
334	0.304998
335	0.36252
336	0.36252
337	0.36252
338	0.36252
339	0.36252
340	0.36252
341	0.36252
342	0.36252
343	0.381688
344	0.381688
345	0.381688
346	0.381688
347	0.387977
348	0.387977
349	0.390419
350	0.391603
351	0.391603
352	0.391603
353	0.391603
354	0.391603
355	0.391603
356	0.391603
357	0.391603
358	0.391603
359	0.395477
360	0.395477
361	0.395477
362	0.395477
363	0.39653
364	0.420122
365	0.420175
366	0.42021
367	0.42021
368	0.42021
369	0.448821
370	0.463262
371	0.463267
372	0.470012
373	0.470014
374	0.470014
375	0.470014
376	0.473503
377	0.473505
378	0.473507
379	0.473507
380	0.473507
381	0.473507
382	0.473507
383	0.473507
384	0.473507
385	0.473507
386	0.473507
387	0.473507
388	0.473507
389	0.473507
390	0.473507
391	0.473507
392	0.473507
393	0.473507
394	0.473507
395	0.473507
396	0.473507
397	0.473507
398	0.473507
399	0.480984
400	0.480984
401	0.480984
402	0.480984
403	0.480984
404	0.480984
405	0.480984
406	0.480984
407	0.482792
408	0.482792
409	0.521421
410	0.540994
411	0.541008
412	0.550557
413	0.550561
414	0.550561
415	0.550561
416	0.550561
417	0.550561
418	0.550561
419	0.550561
420	0.550561
421	0.576468
422	0.589562
423	0.589566
424	0.589566
425	0.599209
426	0.604189
427	0.604189
428	0.606413
429	0.606413
430	0.606413
431	0.606413
432	0.606413
433	0.6098
434	0.611526
435	0.611526
436	0.612307
437	0.612307
438	0.612307
439	0.612307
440	0.612644
441	0.612644
442	0.612644
443	0.612644
444	0.612644
445	0.612644
446	0.612644
447	0.612644
448	0.612644
449	0.612644
450	0.612644
451	0.62518
452	0.62518
453	0.630403
454	0.633017
455	0.633017
456	0.633017
457	0.634938
458	0.635899
459	0.635899
460	0.636328
461	0.636328
462	0.636328
463	0.636328
464	0.636328
465	0.636973
466	0.637255
467	0.637255
468	0.637433
469	0.637433
470	0.637433
471	0.637433
472	0.63749
473	0.63749
474	0.63749
475	0.63749
476	0.63749
477	0.63749
478	0.63749
479	0.63749
480	0.63749
481	0.637668
482	0.637772
483	0.637772
484	0.637799
485	0.637799
486	0.637799
487	0.637799
488	0.637825
489	0.637825
490	0.637825
491	0.637825
492	0.637825
493	0.637825
494	0.637825
495	0.637825
496	0.637838
497	0.637838
498	0.637838
499	0.637838
500	0.637838
501	0.637838
502	0.637838
503	0.637838
504	0.637838
505	0.637838
506	0.637838
507	0.637838
508	0.637838
509	0.637838
510	0.637838
511	0.637838
512	0.637838
513	0.637838
514	0.637838
515	0.637838
516	0.637838
517	0.637838
518	0.637838
519	0.637838
520	0.637838
521	0.637838
522	0.637838
523	0.637838
524	0.637838
525	0.637838
526	0.637838
527	0.637838
528	0.637838
529	0.637838
530	0.637838
531	0.637838
532	0.637838
533	0.637838
534	0.637838
535	0.637838
536	0.637838
537	0.637838
538	0.637838
539	0.637838
540	0.637838
541	0.637838
542	0.637838
543	0.687744
544	0.687744
545	0.687744
546	0.687744
547	0.687744
548	0.687744
549	0.687744
550	0.687744
551	0.687744
552	0.687744
553	0.687744
554	0.687744
555	0.687744
556	0.687744
557	0.687744
558	0.687744
559	0.702128
560	0.702128
561	0.702128
562	0.702128
563	0.702128
564	0.702128
565	0.702128
566	0.702128
567	0.705989
568	0.705989
569	0.705989
570	0.705989
571	0.706991
572	0.728176
573	0.728256
574	0.72828
575	0.72828
576	0.72828
577	0.72828
578	0.72828
579	0.72828
580	0.72828
581	0.72828
582	0.72828
583	0.72828
584	0.72828
585	0.72828
586	0.72828
587	0.72828
588	0.72828
589	0.72828
590	0.72828
591	0.730149
592	0.730149
593	0.730149
594	0.730149
595	0.730149
596	0.730149
597	0.762464
598	0.778691
599	0.778697
600	0.778697
601	0.791967
602	0.798575
603	0.798575
604	0.801342
605	0.801342
606	0.801342
607	0.801342
608	0.801342
609	0.801342
610	0.801342
611	0.820702
612	0.820702
613	0.828613
614	0.832528
615	0.832528
616	0.832528
617	0.835583
618	0.837056
619	0.837056
620	0.837758
621	0.837758
622	0.837758
623	0.837758
624	0.837758
625	0.83867
626	0.839169
627	0.839169
628	0.839357
629	0.839357
630	0.839357
631	0.839357
632	0.839459
633	0.839459
634	0.839459
635	0.839459
636	0.839459
637	0.839459
638	0.839459
639	0.839459
640	0.839459
641	0.839459
642	0.839459
643	0.839459
644	0.839459
645	0.839459
646	0.839459
647	0.839459
648	0.839459
649	0.839459
650	0.839459
651	0.868193
652	0.868193
653	0.8808
654	0.886864
655	0.886864
656	0.886864
657	0.886864
658	0.886864
659	0.89593
660	0.89593
661	0.899558
662	0.901516
663	0.901516
664	0.901516
665	0.902716
666	0.90334
667	0.90334
668	0.903696
669	0.903696
670	0.903696
671	0.903696
672	0.903696
673	0.903696
674	0.903696
675	0.906028
676	0.906028
677	0.906946
678	0.9074
679	0.9074
680	0.9074
681	0.907694
682	0.907809
683	0.907809
684	0.907885
685	0.907885
686	0.907885
687	0.907885
688	0.917262
689	0.917272
690	0.917282
691	0.917282
692	0.917287
693	0.917287
694	0.917287
695	0.917287
696	0.917289
697	0.917289
698	0.917289
699	0.917289
700	0.917289
701	0.917289
702	0.917289
703	0.917289
704	0.917289
705	0.917289
706	0.917289
707	0.917692
708	0.928951
709	0.928975
710	0.928985
711	0.928985
712	0.93425
713	0.934252
714	0.934252
715	0.934252
716	0.934252
717	0.934252
718	0.934252
719	0.934252
720	0.936894
721	0.936896
722	0.936896
723	0.936896
724	0.936896
725	0.936896
726	0.936896
727	0.936896
728	0.936896
729	0.936896
730	0.936896
731	0.936896
732	0.936896
733	0.936896
734	0.936896
735	0.936896
736	0.938187
737	0.938187
738	0.938189
739	0.938189
740	0.938189
741	0.938189
742	0.938189
743	0.938189
744	0.938189
745	0.938189
746	0.938189
747	0.938189
748	0.938189
749	0.938189
750	0.938189
751	0.938189
752	0.938189
753	0.938189
754	0.938189
755	0.938189
756	0.938189
757	0.938189
758	0.938189
759	0.938189
760	0.938189
761	0.938189
762	0.938189
763	0.938189
764	0.938189
765	0.938189
766	0.938189
767	0.938189
768	0.938189
769	0.938189
770	0.938189
771	0.938189
772	0.938189
773	0.938189
774	0.938189
775	0.947495
776	0.947495
777	0.947495
778	0.947495
779	0.950473
780	0.950473
781	0.951638
782	0.952254
783	0.952254
784	0.952254
785	0.952254
786	0.952254
787	0.953031
788	0.953031
789	0.953326
790	0.953444
791	0.953444
792	0.962392
793	0.962402
794	0.962409
795	0.962409
796	0.962413
797	0.962413
798	0.962413
799	0.962413
800	0.962413
801	0.962413
802	0.973473
803	0.9735
804	0.979085
805	0.97909
806	0.97909
807	0.97909
808	0.981724
809	0.981726
810	0.981726
811	0.981726
812	0.981726
813	0.981726
814	0.981726
815	0.981726
816	0.983031
817	0.983031
818	0.983031
819	0.983031
820	0.983031
821	0.983031
822	0.983031
823	0.983031
824	0.983031
825	0.983031
826	0.983031
827	0.983031
828	0.983031
829	0.983031
830	0.983031
831	0.983031
832	0.983031
833	0.989577
834	0.992858
835	0.992858
836	0.99448
837	0.99448
838	0.99448
839	0.99448
840	0.995284
841	0.995284
842	0.995284
843	0.995284
844	0.995284
845	0.995284
846	0.995284
847	0.995284
848	0.99565
849	0.99565
850	0.99565
851	0.99565
852	0.99565
853	0.99565
854	0.99565
855	0.99565
856	0.99565
857	0.99565
858	0.99565
859	0.99565
860	0.99565
861	0.99565
862	0.99565
863	0.99565
864	0.995801
865	0.995801
866	0.995801
867	0.995801
868	0.995801
869	0.995801
870	0.995801
871	0.995801
872	0.995801
873	0.995801
874	0.995801
875	0.995801
876	0.995801
877	0.995801
878	0.995801
879	0.995801
880	0.995801
881	0.995801
882	0.995801
883	0.995801
884	0.995801
885	0.995801
886	0.995801
887	0.995801
888	0.995801
889	0.995801
890	0.995801
891	0.995801
892	0.995801
893	0.995801
894	0.995801
895	0.995801
896	0.995801
897	0.997949
898	0.99901
899	0.99901
900	0.999536
901	0.999536
902	0.999536
903	0.999536
904	0.999789
905	0.999789
906	0.999789
907	0.999789
908	0.999789
909	0.999789
910	0.999789
911	0.999789
912	0.999914
913	0.999914
914	0.999914
915	0.999914
916	0.999914
917	0.999914
918	0.999914
919	0.999914
920	0.999914
921	0.999914
922	0.999914
923	0.999914
924	0.999914
925	0.999914
926	0.999914
927	0.999914
928	0.999975
929	0.999975
930	0.999975
931	0.999975
932	0.999975
933	0.999975
934	0.999975
935	0.999975
936	0.999975
937	0.999975
938	0.999975
939	0.999975
940	0.999975
941	0.999975
942	0.999975
943	0.999975
944	0.999975
945	0.999975
946	0.999975
947	0.999975
948	0.999975
949	0.999975
950	0.999975
951	0.999975
952	0.999975
953	0.999975
954	0.999975
955	0.999975
956	0.999975
957	0.999975
958	0.999975
959	0.999975
960	1
961	1
962	1
963	1
964	1
965	1
966	1
967	1
968	1
969	1
970	1
971	1
972	1
973	1
974	1
975	1
976	1
977	1
978	1
979	1
980	1
981	1
982	1
983	1
984	1
985	1
986	1
987	1
988	1
989	1
990	1
991	1
992	1
993	1
994	1
995	1
996	1
997	1
998	1
999	1
1000	1
1001	1
1002	1
1003	1
1004	1
1005	1
1006	1
1007	1
1008	1
1009	1
1010	1
1011	1
1012	1
1013	1
1014	1
1015	1
1016	1
1017	1
1018	1
1019	1
1020	1
1021	1
1022	1
1023	1
};
\addlegendentry{$E_b/N_0 = 1.5$ dB}

\addplot[
    color=blue,
    thick,
    mark size=3,
]
table {
0	0
1	0
2	0
3	0
4	0
5	0
6	0
7	0
8	0
9	0
10	0
11	0
12	0
13	0
14	0
15	0
16	0
17	0
18	0
19	0
20	0
21	0
22	0
23	0
24	0
25	0
26	0
27	0
28	0
29	0
30	0
31	0
32	0
33	0
34	0
35	0
36	0
37	0
38	0
39	0
40	0
41	0
42	0
43	0
44	0
45	0
46	0
47	0
48	0
49	0
50	0
51	0
52	0
53	0
54	0
55	0
56	0
57	0
58	0
59	0
60	0
61	0
62	0
63	0
64	0
65	0
66	0
67	0
68	0
69	0
70	0
71	0
72	0
73	0
74	0
75	0
76	0
77	0
78	0
79	0
80	0
81	0
82	0
83	0
84	0
85	0
86	0
87	0
88	0
89	0
90	0
91	0
92	0
93	0
94	0
95	0
96	0
97	0
98	0
99	0
100	0
101	0
102	0
103	0
104	0
105	0
106	0
107	0
108	0
109	0
110	0
111	0
112	0
113	0
114	0
115	0
116	0
117	0
118	0
119	0
120	0
121	0
122	0
123	0
124	0
125	0
126	0
127	0.0261579
128	0.0261579
129	0.0261579
130	0.0261579
131	0.0261579
132	0.0261579
133	0.0261579
134	0.0261579
135	0.0261579
136	0.0261579
137	0.0261579
138	0.0261579
139	0.0261579
140	0.0261579
141	0.0261579
142	0.0261579
143	0.0261579
144	0.0261579
145	0.0261579
146	0.0261579
147	0.0261579
148	0.0261579
149	0.0261579
150	0.0261579
151	0.0261579
152	0.0261579
153	0.0261579
154	0.0261579
155	0.0261579
156	0.0261579
157	0.0261579
158	0.0261579
159	0.0261579
160	0.0261579
161	0.0261579
162	0.0261579
163	0.0261579
164	0.0261579
165	0.0261579
166	0.0261579
167	0.0261579
168	0.0261579
169	0.0261579
170	0.0261579
171	0.0261579
172	0.0261579
173	0.0261579
174	0.0261579
175	0.0261579
176	0.0261579
177	0.0261579
178	0.0261579
179	0.0261579
180	0.0261579
181	0.0261579
182	0.0261579
183	0.0261579
184	0.0261579
185	0.0261579
186	0.0261579
187	0.0261579
188	0.0261579
189	0.0261579
190	0.0261579
191	0.028304
192	0.028304
193	0.028304
194	0.028304
195	0.028304
196	0.028304
197	0.028304
198	0.028304
199	0.028304
200	0.028304
201	0.028304
202	0.028304
203	0.028304
204	0.028304
205	0.028304
206	0.028304
207	0.028304
208	0.028304
209	0.028304
210	0.028304
211	0.028304
212	0.028304
213	0.028304
214	0.028304
215	0.028304
216	0.028304
217	0.028304
218	0.028304
219	0.028304
220	0.028304
221	0.0641063
222	0.0815736
223	0.0815736
224	0.0815736
225	0.0815736
226	0.0815736
227	0.0815736
228	0.0815736
229	0.0815736
230	0.0815736
231	0.129571
232	0.129571
233	0.129571
234	0.129571
235	0.145004
236	0.145004
237	0.150482
238	0.153436
239	0.153436
240	0.153436
241	0.153436
242	0.153436
243	0.156828
244	0.156828
245	0.157994
246	0.158597
247	0.158597
248	0.158597
249	0.159047
250	0.159272
251	0.159272
252	0.159372
253	0.159372
254	0.159372
255	0.159372
256	0.159372
257	0.159372
258	0.159372
259	0.159372
260	0.159372
261	0.159372
262	0.159372
263	0.159372
264	0.159372
265	0.159372
266	0.159372
267	0.159372
268	0.159372
269	0.159372
270	0.159372
271	0.159372
272	0.159372
273	0.159372
274	0.159372
275	0.159372
276	0.159372
277	0.159372
278	0.159372
279	0.159372
280	0.159372
281	0.159372
282	0.159372
283	0.159372
284	0.159372
285	0.159372
286	0.159372
287	0.159372
288	0.159372
289	0.159372
290	0.159372
291	0.159372
292	0.159372
293	0.159372
294	0.159372
295	0.159372
296	0.159372
297	0.159372
298	0.159372
299	0.159372
300	0.159372
301	0.159372
302	0.159372
303	0.159372
304	0.159372
305	0.159372
306	0.159372
307	0.159372
308	0.159372
309	0.159372
310	0.159372
311	0.159372
312	0.159372
313	0.159372
314	0.159372
315	0.193902
316	0.193902
317	0.207362
318	0.21388
319	0.21388
320	0.21388
321	0.21388
322	0.21388
323	0.21388
324	0.21388
325	0.21388
326	0.21388
327	0.21388
328	0.21388
329	0.21388
330	0.21388
331	0.21388
332	0.21388
333	0.21388
334	0.21388
335	0.264997
336	0.264997
337	0.264997
338	0.264997
339	0.264997
340	0.264997
341	0.264997
342	0.264997
343	0.277112
344	0.277112
345	0.277112
346	0.277112
347	0.280662
348	0.280662
349	0.281722
350	0.282364
351	0.282364
352	0.282364
353	0.282364
354	0.282364
355	0.282364
356	0.282364
357	0.282364
358	0.282364
359	0.28402
360	0.28402
361	0.28402
362	0.28402
363	0.284398
364	0.309052
365	0.309052
366	0.309052
367	0.309052
368	0.309052
369	0.340026
370	0.355175
371	0.355175
372	0.362547
373	0.362547
374	0.362547
375	0.362547
376	0.366296
377	0.366296
378	0.366296
379	0.366296
380	0.366296
381	0.366296
382	0.366296
383	0.366296
384	0.366296
385	0.366296
386	0.366296
387	0.366296
388	0.366296
389	0.366296
390	0.366296
391	0.366296
392	0.366296
393	0.366296
394	0.366296
395	0.366296
396	0.366296
397	0.366296
398	0.366296
399	0.370257
400	0.370257
401	0.370257
402	0.370257
403	0.370257
404	0.370257
405	0.370257
406	0.370257
407	0.370926
408	0.370926
409	0.413306
410	0.434536
411	0.434542
412	0.444988
413	0.444988
414	0.444988
415	0.444988
416	0.444988
417	0.444988
418	0.444988
419	0.444988
420	0.444988
421	0.472544
422	0.485845
423	0.485845
424	0.485845
425	0.495469
426	0.500623
427	0.500623
428	0.502915
429	0.502915
430	0.502915
431	0.502915
432	0.502915
433	0.506054
434	0.507386
435	0.507386
436	0.507975
437	0.507975
438	0.507975
439	0.507975
440	0.508313
441	0.508313
442	0.508313
443	0.508313
444	0.508313
445	0.508313
446	0.508313
447	0.508313
448	0.508313
449	0.508313
450	0.508313
451	0.520792
452	0.520792
453	0.525999
454	0.528635
455	0.528635
456	0.528635
457	0.530066
458	0.530894
459	0.530894
460	0.531305
461	0.531305
462	0.531305
463	0.531305
464	0.531305
465	0.531768
466	0.532033
467	0.532033
468	0.532119
469	0.532119
470	0.532119
471	0.532119
472	0.532166
473	0.532166
474	0.532166
475	0.532166
476	0.532166
477	0.532166
478	0.532166
479	0.532166
480	0.532166
481	0.532318
482	0.532358
483	0.532358
484	0.532411
485	0.532411
486	0.532411
487	0.532411
488	0.532437
489	0.532437
490	0.532437
491	0.532437
492	0.532437
493	0.532437
494	0.532437
495	0.532437
496	0.532444
497	0.532444
498	0.532444
499	0.532444
500	0.532444
501	0.532444
502	0.532444
503	0.532444
504	0.532444
505	0.532444
506	0.532444
507	0.532444
508	0.532444
509	0.532444
510	0.532444
511	0.532444
512	0.532444
513	0.532444
514	0.532444
515	0.532444
516	0.532444
517	0.532444
518	0.532444
519	0.532444
520	0.532444
521	0.532444
522	0.532444
523	0.532444
524	0.532444
525	0.532444
526	0.532444
527	0.532444
528	0.532444
529	0.532444
530	0.532444
531	0.532444
532	0.532444
533	0.532444
534	0.532444
535	0.532444
536	0.532444
537	0.532444
538	0.532444
539	0.532444
540	0.532444
541	0.532444
542	0.532444
543	0.590039
544	0.590039
545	0.590039
546	0.590039
547	0.590039
548	0.590039
549	0.590039
550	0.590039
551	0.590039
552	0.590039
553	0.590039
554	0.590039
555	0.590039
556	0.590039
557	0.590039
558	0.590039
559	0.601266
560	0.601266
561	0.601266
562	0.601266
563	0.601266
564	0.601266
565	0.601266
566	0.601266
567	0.603399
568	0.603399
569	0.603399
570	0.603399
571	0.603843
572	0.629908
573	0.629915
574	0.629922
575	0.629922
576	0.629922
577	0.629922
578	0.629922
579	0.629922
580	0.629922
581	0.629922
582	0.629922
583	0.629922
584	0.629922
585	0.629922
586	0.629922
587	0.629922
588	0.629922
589	0.629922
590	0.629922
591	0.630796
592	0.630796
593	0.630796
594	0.630796
595	0.630796
596	0.630796
597	0.670738
598	0.690338
599	0.690338
600	0.690338
601	0.705342
602	0.712913
603	0.712913
604	0.716132
605	0.716132
606	0.716132
607	0.716132
608	0.716132
609	0.716132
610	0.716132
611	0.738978
612	0.738978
613	0.747529
614	0.751934
615	0.751934
616	0.751934
617	0.754882
618	0.756313
619	0.756313
620	0.756995
621	0.756995
622	0.756995
623	0.756995
624	0.756995
625	0.758008
626	0.758472
627	0.758472
628	0.75875
629	0.75875
630	0.75875
631	0.75875
632	0.758869
633	0.758869
634	0.758869
635	0.758869
636	0.758869
637	0.758869
638	0.758869
639	0.758869
640	0.758869
641	0.758869
642	0.758869
643	0.758869
644	0.758869
645	0.758869
646	0.758869
647	0.758869
648	0.758869
649	0.758869
650	0.758869
651	0.797156
652	0.797156
653	0.812556
654	0.820717
655	0.820717
656	0.820717
657	0.820717
658	0.820717
659	0.830858
660	0.830858
661	0.834541
662	0.836436
663	0.836436
664	0.836436
665	0.837621
666	0.838178
667	0.838178
668	0.838555
669	0.838555
670	0.838555
671	0.838555
672	0.838555
673	0.838555
674	0.838555
675	0.840681
676	0.840681
677	0.841397
678	0.841808
679	0.841808
680	0.841808
681	0.842006
682	0.842072
683	0.842072
684	0.842125
685	0.842125
686	0.842125
687	0.842125
688	0.85893
689	0.858944
690	0.858944
691	0.858944
692	0.85895
693	0.85895
694	0.85895
695	0.85895
696	0.85895
697	0.85895
698	0.85895
699	0.85895
700	0.85895
701	0.85895
702	0.85895
703	0.85895
704	0.85895
705	0.85895
706	0.85895
707	0.859156
708	0.87961
709	0.879617
710	0.879617
711	0.879617
712	0.889838
713	0.889838
714	0.889838
715	0.889838
716	0.889838
717	0.889838
718	0.889838
719	0.889838
720	0.894766
721	0.894766
722	0.894766
723	0.894766
724	0.894766
725	0.894766
726	0.894766
727	0.894766
728	0.894766
729	0.894766
730	0.894766
731	0.894766
732	0.894766
733	0.894766
734	0.894766
735	0.894766
736	0.897429
737	0.897429
738	0.897429
739	0.897429
740	0.897429
741	0.897429
742	0.897429
743	0.897429
744	0.897429
745	0.897429
746	0.897429
747	0.897429
748	0.897429
749	0.897429
750	0.897429
751	0.897429
752	0.897429
753	0.897429
754	0.897429
755	0.897429
756	0.897429
757	0.897429
758	0.897429
759	0.897429
760	0.897429
761	0.897429
762	0.897429
763	0.897429
764	0.897429
765	0.897429
766	0.897429
767	0.897429
768	0.897429
769	0.897429
770	0.897429
771	0.897429
772	0.897429
773	0.897429
774	0.897429
775	0.908404
776	0.908404
777	0.908404
778	0.908404
779	0.911511
780	0.911511
781	0.912677
782	0.91322
783	0.91322
784	0.91322
785	0.91322
786	0.91322
787	0.913849
788	0.913849
789	0.914055
790	0.914108
791	0.914108
792	0.930383
793	0.930396
794	0.930402
795	0.930402
796	0.930402
797	0.930402
798	0.930402
799	0.930402
800	0.930402
801	0.930402
802	0.951645
803	0.951645
804	0.961886
805	0.961893
806	0.961893
807	0.961893
808	0.966967
809	0.966967
810	0.966967
811	0.966967
812	0.966967
813	0.966967
814	0.966967
815	0.966967
816	0.969378
817	0.969378
818	0.969378
819	0.969378
820	0.969378
821	0.969378
822	0.969378
823	0.969378
824	0.969378
825	0.969378
826	0.969378
827	0.969378
828	0.969378
829	0.969378
830	0.969378
831	0.969378
832	0.969378
833	0.98146
834	0.987554
835	0.987554
836	0.990276
837	0.990276
838	0.990276
839	0.990276
840	0.991647
841	0.991647
842	0.991647
843	0.991647
844	0.991647
845	0.991647
846	0.991647
847	0.991647
848	0.992224
849	0.992224
850	0.992224
851	0.992224
852	0.992224
853	0.992224
854	0.992224
855	0.992224
856	0.992224
857	0.992224
858	0.992224
859	0.992224
860	0.992224
861	0.992224
862	0.992224
863	0.992224
864	0.992508
865	0.992508
866	0.992508
867	0.992508
868	0.992508
869	0.992508
870	0.992508
871	0.992508
872	0.992508
873	0.992508
874	0.992508
875	0.992508
876	0.992508
877	0.992508
878	0.992508
879	0.992508
880	0.992508
881	0.992508
882	0.992508
883	0.992508
884	0.992508
885	0.992508
886	0.992508
887	0.992508
888	0.992508
889	0.992508
890	0.992508
891	0.992508
892	0.992508
893	0.992508
894	0.992508
895	0.992508
896	0.992508
897	0.996363
898	0.998437
899	0.998437
900	0.999245
901	0.999245
902	0.999245
903	0.999245
904	0.999622
905	0.999622
906	0.999622
907	0.999622
908	0.999622
909	0.999622
910	0.999622
911	0.999622
912	0.999808
913	0.999808
914	0.999808
915	0.999808
916	0.999808
917	0.999808
918	0.999808
919	0.999808
920	0.999808
921	0.999808
922	0.999808
923	0.999808
924	0.999808
925	0.999808
926	0.999808
927	0.999808
928	0.99994
929	0.99994
930	0.99994
931	0.99994
932	0.99994
933	0.99994
934	0.99994
935	0.99994
936	0.99994
937	0.99994
938	0.99994
939	0.99994
940	0.99994
941	0.99994
942	0.99994
943	0.99994
944	0.99994
945	0.99994
946	0.99994
947	0.99994
948	0.99994
949	0.99994
950	0.99994
951	0.99994
952	0.99994
953	0.99994
954	0.99994
955	0.99994
956	0.99994
957	0.99994
958	0.99994
959	0.99994
960	1
961	1
962	1
963	1
964	1
965	1
966	1
967	1
968	1
969	1
970	1
971	1
972	1
973	1
974	1
975	1
976	1
977	1
978	1
979	1
980	1
981	1
982	1
983	1
984	1
985	1
986	1
987	1
988	1
989	1
990	1
991	1
992	1
993	1
994	1
995	1
996	1
997	1
998	1
999	1
1000	1
1001	1
1002	1
1003	1
1004	1
1005	1
1006	1
1007	1
1008	1
1009	1
1010	1
1011	1
1012	1
1013	1
1014	1
1015	1
1016	1
1017	1
1018	1
1019	1
1020	1
1021	1
1022	1
1023	1
};
\addlegendentry{$E_b/N_0 = 2.0$ dB}

\addplot[
    color=red,
    dashed,
    thick,
    mark size=3,
]
table {
0	0
1	0
2	0
3	0
4	0
5	0
6	0
7	0
8	0
9	0
10	0
11	0
12	0
13	0
14	0
15	0
16	0
17	0
18	0
19	0
20	0
21	0
22	0
23	0
24	0
25	0
26	0
27	0
28	0
29	0
30	0
31	0
32	0
33	0
34	0
35	0
36	0
37	0
38	0
39	0
40	0
41	0
42	0
43	0
44	0
45	0
46	0
47	0
48	0
49	0
50	0
51	0
52	0
53	0
54	0
55	0
56	0
57	0
58	0
59	0
60	0
61	0
62	0
63	0
64	0
65	0
66	0
67	0
68	0
69	0
70	0
71	0
72	0
73	0
74	0
75	0
76	0
77	0
78	0
79	0
80	0
81	0
82	0
83	0
84	0
85	0
86	0
87	0
88	0
89	0
90	0
91	0
92	0
93	0
94	0
95	0
96	0
97	0
98	0
99	0
100	0
101	0
102	0
103	0
104	0
105	0
106	0
107	0
108	0
109	0
110	0
111	0
112	0
113	0
114	0
115	0
116	0
117	0
118	0
119	0
120	0
121	0
122	0
123	0
124	0
125	0
126	0
127	0.0163922
128	0.0163922
129	0.0163922
130	0.0163922
131	0.0163922
132	0.0163922
133	0.0163922
134	0.0163922
135	0.0163922
136	0.0163922
137	0.0163922
138	0.0163922
139	0.0163922
140	0.0163922
141	0.0163922
142	0.0163922
143	0.0163922
144	0.0163922
145	0.0163922
146	0.0163922
147	0.0163922
148	0.0163922
149	0.0163922
150	0.0163922
151	0.0163922
152	0.0163922
153	0.0163922
154	0.0163922
155	0.0163922
156	0.0163922
157	0.0163922
158	0.0163922
159	0.0163922
160	0.0163922
161	0.0163922
162	0.0163922
163	0.0163922
164	0.0163922
165	0.0163922
166	0.0163922
167	0.0163922
168	0.0163922
169	0.0163922
170	0.0163922
171	0.0163922
172	0.0163922
173	0.0163922
174	0.0163922
175	0.0163922
176	0.0163922
177	0.0163922
178	0.0163922
179	0.0163922
180	0.0163922
181	0.0163922
182	0.0163922
183	0.0163922
184	0.0163922
185	0.0163922
186	0.0163922
187	0.0163922
188	0.0163922
189	0.0163922
190	0.0163922
191	0.0168594
192	0.0168594
193	0.0168594
194	0.0168594
195	0.0168594
196	0.0168594
197	0.0168594
198	0.0168594
199	0.0168594
200	0.0168594
201	0.0168594
202	0.0168594
203	0.0168594
204	0.0168594
205	0.0168594
206	0.0168594
207	0.0168594
208	0.0168594
209	0.0168594
210	0.0168594
211	0.0168594
212	0.0168594
213	0.0168594
214	0.0168594
215	0.0168594
216	0.0168594
217	0.0168594
218	0.0168594
219	0.0168594
220	0.0168594
221	0.0419733
222	0.0545108
223	0.0545108
224	0.0545108
225	0.0545108
226	0.0545108
227	0.0545108
228	0.0545108
229	0.0545108
230	0.0545108
231	0.0890472
232	0.0890472
233	0.0890472
234	0.0890472
235	0.0988592
236	0.0988592
237	0.101779
238	0.103181
239	0.103181
240	0.103181
241	0.103181
242	0.103181
243	0.105128
244	0.105128
245	0.105673
246	0.106062
247	0.106062
248	0.106062
249	0.106218
250	0.106218
251	0.106218
252	0.106218
253	0.106218
254	0.106218
255	0.106218
256	0.106218
257	0.106218
258	0.106218
259	0.106218
260	0.106218
261	0.106218
262	0.106218
263	0.106218
264	0.106218
265	0.106218
266	0.106218
267	0.106218
268	0.106218
269	0.106218
270	0.106218
271	0.106218
272	0.106218
273	0.106218
274	0.106218
275	0.106218
276	0.106218
277	0.106218
278	0.106218
279	0.106218
280	0.106218
281	0.106218
282	0.106218
283	0.106218
284	0.106218
285	0.106218
286	0.106218
287	0.106218
288	0.106218
289	0.106218
290	0.106218
291	0.106218
292	0.106218
293	0.106218
294	0.106218
295	0.106218
296	0.106218
297	0.106218
298	0.106218
299	0.106218
300	0.106218
301	0.106218
302	0.106218
303	0.106218
304	0.106218
305	0.106218
306	0.106218
307	0.106218
308	0.106218
309	0.106218
310	0.106218
311	0.106218
312	0.106218
313	0.106218
314	0.106218
315	0.128762
316	0.128762
317	0.137095
318	0.140794
319	0.140794
320	0.140794
321	0.140794
322	0.140794
323	0.140794
324	0.140794
325	0.140794
326	0.140794
327	0.140794
328	0.140794
329	0.140794
330	0.140794
331	0.140794
332	0.140794
333	0.140794
334	0.140794
335	0.179457
336	0.179457
337	0.179457
338	0.179457
339	0.179457
340	0.179457
341	0.179457
342	0.179457
343	0.186388
344	0.186388
345	0.186388
346	0.186388
347	0.1874
348	0.1874
349	0.188179
350	0.188374
351	0.188374
352	0.188374
353	0.188374
354	0.188374
355	0.188374
356	0.188374
357	0.188374
358	0.188374
359	0.188919
360	0.188919
361	0.188919
362	0.188919
363	0.189074
364	0.217498
365	0.217498
366	0.217498
367	0.217498
368	0.217498
369	0.249075
370	0.265467
371	0.265467
372	0.273605
373	0.273605
374	0.273605
375	0.273605
376	0.27781
377	0.27781
378	0.27781
379	0.27781
380	0.27781
381	0.27781
382	0.27781
383	0.27781
384	0.27781
385	0.27781
386	0.27781
387	0.27781
388	0.27781
389	0.27781
390	0.27781
391	0.27781
392	0.27781
393	0.27781
394	0.27781
395	0.27781
396	0.27781
397	0.27781
398	0.27781
399	0.279368
400	0.279368
401	0.279368
402	0.279368
403	0.279368
404	0.279368
405	0.279368
406	0.279368
407	0.27964
408	0.27964
409	0.323872
410	0.347273
411	0.347273
412	0.359499
413	0.359499
414	0.359499
415	0.359499
416	0.359499
417	0.359499
418	0.359499
419	0.359499
420	0.359499
421	0.386326
422	0.39898
423	0.39898
424	0.39898
425	0.408208
426	0.412257
427	0.412257
428	0.414165
429	0.414165
430	0.414165
431	0.414165
432	0.414165
433	0.416968
434	0.418487
435	0.418487
436	0.419227
437	0.419227
438	0.419227
439	0.419227
440	0.419538
441	0.419538
442	0.419538
443	0.419538
444	0.419538
445	0.419538
446	0.419538
447	0.419538
448	0.419538
449	0.419538
450	0.419538
451	0.430674
452	0.430674
453	0.434529
454	0.437293
455	0.437293
456	0.437293
457	0.438461
458	0.439045
459	0.439045
460	0.439435
461	0.439435
462	0.439435
463	0.439435
464	0.439435
465	0.43998
466	0.440135
467	0.440135
468	0.440369
469	0.440369
470	0.440369
471	0.440369
472	0.440447
473	0.440447
474	0.440447
475	0.440447
476	0.440447
477	0.440447
478	0.440447
479	0.440447
480	0.440447
481	0.440486
482	0.440486
483	0.440486
484	0.440525
485	0.440525
486	0.440525
487	0.440525
488	0.440525
489	0.440525
490	0.440525
491	0.440525
492	0.440525
493	0.440525
494	0.440525
495	0.440525
496	0.440525
497	0.440525
498	0.440525
499	0.440525
500	0.440525
501	0.440525
502	0.440525
503	0.440525
504	0.440525
505	0.440525
506	0.440525
507	0.440525
508	0.440525
509	0.440525
510	0.440525
511	0.440525
512	0.440525
513	0.440525
514	0.440525
515	0.440525
516	0.440525
517	0.440525
518	0.440525
519	0.440525
520	0.440525
521	0.440525
522	0.440525
523	0.440525
524	0.440525
525	0.440525
526	0.440525
527	0.440525
528	0.440525
529	0.440525
530	0.440525
531	0.440525
532	0.440525
533	0.440525
534	0.440525
535	0.440525
536	0.440525
537	0.440525
538	0.440525
539	0.440525
540	0.440525
541	0.440525
542	0.440525
543	0.491025
544	0.491025
545	0.491025
546	0.491025
547	0.491025
548	0.491025
549	0.491025
550	0.491025
551	0.491025
552	0.491025
553	0.491025
554	0.491025
555	0.491025
556	0.491025
557	0.491025
558	0.491025
559	0.496905
560	0.496905
561	0.496905
562	0.496905
563	0.496905
564	0.496905
565	0.496905
566	0.496905
567	0.497683
568	0.497683
569	0.497683
570	0.497683
571	0.497761
572	0.527119
573	0.527119
574	0.527119
575	0.527119
576	0.527119
577	0.527119
578	0.527119
579	0.527119
580	0.527119
581	0.527119
582	0.527119
583	0.527119
584	0.527119
585	0.527119
586	0.527119
587	0.527119
588	0.527119
589	0.527119
590	0.527119
591	0.527431
592	0.527431
593	0.527431
594	0.527431
595	0.527431
596	0.527431
597	0.569482
598	0.591247
599	0.591247
600	0.591247
601	0.605615
602	0.611494
603	0.611494
604	0.615193
605	0.615193
606	0.615193
607	0.615193
608	0.615193
609	0.615193
610	0.615193
611	0.637426
612	0.637426
613	0.645485
614	0.649652
615	0.649652
616	0.649652
617	0.652572
618	0.653779
619	0.653779
620	0.654012
621	0.654012
622	0.654012
623	0.654012
624	0.654012
625	0.654713
626	0.654947
627	0.654947
628	0.655025
629	0.655025
630	0.655025
631	0.655025
632	0.655025
633	0.655025
634	0.655025
635	0.655025
636	0.655025
637	0.655025
638	0.655025
639	0.655025
640	0.655025
641	0.655025
642	0.655025
643	0.655025
644	0.655025
645	0.655025
646	0.655025
647	0.655025
648	0.655025
649	0.655025
650	0.655025
651	0.694078
652	0.694078
653	0.708484
654	0.716778
655	0.716778
656	0.716778
657	0.716778
658	0.716778
659	0.726083
660	0.726083
661	0.729276
662	0.731573
663	0.731573
664	0.731573
665	0.732547
666	0.732936
667	0.732936
668	0.733131
669	0.733131
670	0.733131
671	0.733131
672	0.733131
673	0.733131
674	0.733131
675	0.734416
676	0.734416
677	0.734805
678	0.735078
679	0.735078
680	0.735078
681	0.735233
682	0.735311
683	0.735311
684	0.73535
685	0.73535
686	0.73535
687	0.73535
688	0.767044
689	0.767044
690	0.767083
691	0.767083
692	0.767083
693	0.767083
694	0.767083
695	0.767083
696	0.767083
697	0.767083
698	0.767083
699	0.767083
700	0.767083
701	0.767083
702	0.767083
703	0.767083
704	0.767083
705	0.767083
706	0.767083
707	0.7672
708	0.80419
709	0.80419
710	0.80419
711	0.80419
712	0.822996
713	0.822996
714	0.822996
715	0.822996
716	0.822996
717	0.822996
718	0.822996
719	0.822996
720	0.832457
721	0.832457
722	0.832457
723	0.832457
724	0.832457
725	0.832457
726	0.832457
727	0.832457
728	0.832457
729	0.832457
730	0.832457
731	0.832457
732	0.832457
733	0.832457
734	0.832457
735	0.832457
736	0.837013
737	0.837013
738	0.837013
739	0.837013
740	0.837013
741	0.837013
742	0.837013
743	0.837013
744	0.837013
745	0.837013
746	0.837013
747	0.837013
748	0.837013
749	0.837013
750	0.837013
751	0.837013
752	0.837013
753	0.837013
754	0.837013
755	0.837013
756	0.837013
757	0.837013
758	0.837013
759	0.837013
760	0.837013
761	0.837013
762	0.837013
763	0.837013
764	0.837013
765	0.837013
766	0.837013
767	0.837013
768	0.837013
769	0.837013
770	0.837013
771	0.837013
772	0.837013
773	0.837013
774	0.837013
775	0.847448
776	0.847448
777	0.847448
778	0.847448
779	0.850018
780	0.850018
781	0.850602
782	0.850835
783	0.850835
784	0.850835
785	0.850835
786	0.850835
787	0.85138
788	0.85138
789	0.851497
790	0.851614
791	0.851614
792	0.881089
793	0.881128
794	0.881128
795	0.881128
796	0.881128
797	0.881128
798	0.881128
799	0.881128
800	0.881128
801	0.881128
802	0.916131
803	0.916131
804	0.932523
805	0.932523
806	0.932523
807	0.932523
808	0.940934
809	0.940934
810	0.940934
811	0.940934
812	0.940934
813	0.940934
814	0.940934
815	0.940934
816	0.945567
817	0.945567
818	0.945567
819	0.945567
820	0.945567
821	0.945567
822	0.945567
823	0.945567
824	0.945567
825	0.945567
826	0.945567
827	0.945567
828	0.945567
829	0.945567
830	0.945567
831	0.945567
832	0.945567
833	0.967371
834	0.977339
835	0.977339
836	0.982206
837	0.982206
838	0.982206
839	0.982206
840	0.985438
841	0.985438
842	0.985438
843	0.985438
844	0.985438
845	0.985438
846	0.985438
847	0.985438
848	0.98684
849	0.98684
850	0.98684
851	0.98684
852	0.98684
853	0.98684
854	0.98684
855	0.98684
856	0.98684
857	0.98684
858	0.98684
859	0.98684
860	0.98684
861	0.98684
862	0.98684
863	0.98684
864	0.987385
865	0.987385
866	0.987385
867	0.987385
868	0.987385
869	0.987385
870	0.987385
871	0.987385
872	0.987385
873	0.987385
874	0.987385
875	0.987385
876	0.987385
877	0.987385
878	0.987385
879	0.987385
880	0.987385
881	0.987385
882	0.987385
883	0.987385
884	0.987385
885	0.987385
886	0.987385
887	0.987385
888	0.987385
889	0.987385
890	0.987385
891	0.987385
892	0.987385
893	0.987385
894	0.987385
895	0.987385
896	0.987385
897	0.994043
898	0.99743
899	0.99743
900	0.998871
901	0.998871
902	0.998871
903	0.998871
904	0.999494
905	0.999494
906	0.999494
907	0.999494
908	0.999494
909	0.999494
910	0.999494
911	0.999494
912	0.999689
913	0.999689
914	0.999689
915	0.999689
916	0.999689
917	0.999689
918	0.999689
919	0.999689
920	0.999689
921	0.999689
922	0.999689
923	0.999689
924	0.999689
925	0.999689
926	0.999689
927	0.999689
928	0.999883
929	0.999883
930	0.999883
931	0.999883
932	0.999883
933	0.999883
934	0.999883
935	0.999883
936	0.999883
937	0.999883
938	0.999883
939	0.999883
940	0.999883
941	0.999883
942	0.999883
943	0.999883
944	0.999883
945	0.999883
946	0.999883
947	0.999883
948	0.999883
949	0.999883
950	0.999883
951	0.999883
952	0.999883
953	0.999883
954	0.999883
955	0.999883
956	0.999883
957	0.999883
958	0.999883
959	0.999883
960	1
961	1
962	1
963	1
964	1
965	1
966	1
967	1
968	1
969	1
970	1
971	1
972	1
973	1
974	1
975	1
976	1
977	1
978	1
979	1
980	1
981	1
982	1
983	1
984	1
985	1
986	1
987	1
988	1
989	1
990	1
991	1
992	1
993	1
994	1
995	1
996	1
997	1
998	1
999	1
1000	1
1001	1
1002	1
1003	1
1004	1
1005	1
1006	1
1007	1
1008	1
1009	1
1010	1
1011	1
1012	1
1013	1
1014	1
1015	1
1016	1
1017	1
1018	1
1019	1
1020	1
1021	1
1022	1
1023	1
};
\addlegendentry{$E_b/N_0 = 2.5$ dB}

\addplot[
    color=black,
    dash pattern=on 3pt off 1pt,
    thick,
    mark size=3,
]
table {
0	0
1	0
2	0
3	0
4	0
5	0
6	0
7	0
8	0
9	0
10	0
11	0
12	0
13	0
14	0
15	0
16	0
17	0
18	0
19	0
20	0
21	0
22	0
23	0
24	0
25	0
26	0
27	0
28	0
29	0
30	0
31	0
32	0
33	0
34	0
35	0
36	0
37	0
38	0
39	0
40	0
41	0
42	0
43	0
44	0
45	0
46	0
47	0
48	0
49	0
50	0
51	0
52	0
53	0
54	0
55	0
56	0
57	0
58	0
59	0
60	0
61	0
62	0
63	0
64	0
65	0
66	0
67	0
68	0
69	0
70	0
71	0
72	0
73	0
74	0
75	0
76	0
77	0
78	0
79	0
80	0
81	0
82	0
83	0
84	0
85	0
86	0
87	0
88	0
89	0
90	0
91	0
92	0
93	0
94	0
95	0
96	0
97	0
98	0
99	0
100	0
101	0
102	0
103	0
104	0
105	0
106	0
107	0
108	0
109	0
110	0
111	0
112	0
113	0
114	0
115	0
116	0
117	0
118	0
119	0
120	0
121	0
122	0
123	0
124	0
125	0
126	0
127	0.00683995
128	0.00683995
129	0.00683995
130	0.00683995
131	0.00683995
132	0.00683995
133	0.00683995
134	0.00683995
135	0.00683995
136	0.00683995
137	0.00683995
138	0.00683995
139	0.00683995
140	0.00683995
141	0.00683995
142	0.00683995
143	0.00683995
144	0.00683995
145	0.00683995
146	0.00683995
147	0.00683995
148	0.00683995
149	0.00683995
150	0.00683995
151	0.00683995
152	0.00683995
153	0.00683995
154	0.00683995
155	0.00683995
156	0.00683995
157	0.00683995
158	0.00683995
159	0.00683995
160	0.00683995
161	0.00683995
162	0.00683995
163	0.00683995
164	0.00683995
165	0.00683995
166	0.00683995
167	0.00683995
168	0.00683995
169	0.00683995
170	0.00683995
171	0.00683995
172	0.00683995
173	0.00683995
174	0.00683995
175	0.00683995
176	0.00683995
177	0.00683995
178	0.00683995
179	0.00683995
180	0.00683995
181	0.00683995
182	0.00683995
183	0.00683995
184	0.00683995
185	0.00683995
186	0.00683995
187	0.00683995
188	0.00683995
189	0.00683995
190	0.00683995
191	0.00683995
192	0.00683995
193	0.00683995
194	0.00683995
195	0.00683995
196	0.00683995
197	0.00683995
198	0.00683995
199	0.00683995
200	0.00683995
201	0.00683995
202	0.00683995
203	0.00683995
204	0.00683995
205	0.00683995
206	0.00683995
207	0.00683995
208	0.00683995
209	0.00683995
210	0.00683995
211	0.00683995
212	0.00683995
213	0.00683995
214	0.00683995
215	0.00683995
216	0.00683995
217	0.00683995
218	0.00683995
219	0.00683995
220	0.00683995
221	0.0212038
222	0.0280438
223	0.0280438
224	0.0280438
225	0.0280438
226	0.0280438
227	0.0280438
228	0.0280438
229	0.0280438
230	0.0280438
231	0.0461696
232	0.0461696
233	0.0461696
234	0.0461696
235	0.0499316
236	0.0499316
237	0.0509576
238	0.0519836
239	0.0519836
240	0.0519836
241	0.0519836
242	0.0519836
243	0.0523256
244	0.0523256
245	0.0523256
246	0.0523256
247	0.0523256
248	0.0523256
249	0.0526676
250	0.0526676
251	0.0526676
252	0.0526676
253	0.0526676
254	0.0526676
255	0.0526676
256	0.0526676
257	0.0526676
258	0.0526676
259	0.0526676
260	0.0526676
261	0.0526676
262	0.0526676
263	0.0526676
264	0.0526676
265	0.0526676
266	0.0526676
267	0.0526676
268	0.0526676
269	0.0526676
270	0.0526676
271	0.0526676
272	0.0526676
273	0.0526676
274	0.0526676
275	0.0526676
276	0.0526676
277	0.0526676
278	0.0526676
279	0.0526676
280	0.0526676
281	0.0526676
282	0.0526676
283	0.0526676
284	0.0526676
285	0.0526676
286	0.0526676
287	0.0526676
288	0.0526676
289	0.0526676
290	0.0526676
291	0.0526676
292	0.0526676
293	0.0526676
294	0.0526676
295	0.0526676
296	0.0526676
297	0.0526676
298	0.0526676
299	0.0526676
300	0.0526676
301	0.0526676
302	0.0526676
303	0.0526676
304	0.0526676
305	0.0526676
306	0.0526676
307	0.0526676
308	0.0526676
309	0.0526676
310	0.0526676
311	0.0526676
312	0.0526676
313	0.0526676
314	0.0526676
315	0.0598495
316	0.0598495
317	0.0639535
318	0.0646375
319	0.0646375
320	0.0646375
321	0.0646375
322	0.0646375
323	0.0646375
324	0.0646375
325	0.0646375
326	0.0646375
327	0.0646375
328	0.0646375
329	0.0646375
330	0.0646375
331	0.0646375
332	0.0646375
333	0.0646375
334	0.0646375
335	0.0865253
336	0.0865253
337	0.0865253
338	0.0865253
339	0.0865253
340	0.0865253
341	0.0865253
342	0.0865253
343	0.0882353
344	0.0882353
345	0.0882353
346	0.0882353
347	0.0889193
348	0.0889193
349	0.0896033
350	0.0896033
351	0.0896033
352	0.0896033
353	0.0896033
354	0.0896033
355	0.0896033
356	0.0896033
357	0.0896033
358	0.0896033
359	0.0896033
360	0.0896033
361	0.0896033
362	0.0896033
363	0.0896033
364	0.114911
365	0.114911
366	0.114911
367	0.114911
368	0.114911
369	0.147743
370	0.164159
371	0.164159
372	0.171341
373	0.171341
374	0.171341
375	0.171341
376	0.173051
377	0.173051
378	0.173051
379	0.173051
380	0.173051
381	0.173051
382	0.173051
383	0.173051
384	0.173051
385	0.173051
386	0.173051
387	0.173051
388	0.173051
389	0.173051
390	0.173051
391	0.173051
392	0.173051
393	0.173051
394	0.173051
395	0.173051
396	0.173051
397	0.173051
398	0.173051
399	0.173735
400	0.173735
401	0.173735
402	0.173735
403	0.173735
404	0.173735
405	0.173735
406	0.173735
407	0.173735
408	0.173735
409	0.214432
410	0.237688
411	0.237688
412	0.248632
413	0.248632
414	0.248632
415	0.248632
416	0.248632
417	0.248632
418	0.248632
419	0.248632
420	0.248632
421	0.27223
422	0.27907
423	0.27907
424	0.27907
425	0.285226
426	0.288304
427	0.288304
428	0.29104
429	0.29104
430	0.29104
431	0.29104
432	0.29104
433	0.294802
434	0.296512
435	0.296512
436	0.297196
437	0.297196
438	0.297196
439	0.297196
440	0.297196
441	0.297196
442	0.297196
443	0.297196
444	0.297196
445	0.297196
446	0.297196
447	0.297196
448	0.297196
449	0.297196
450	0.297196
451	0.306088
452	0.306088
453	0.310534
454	0.311218
455	0.311218
456	0.311218
457	0.311218
458	0.31156
459	0.31156
460	0.31156
461	0.31156
462	0.31156
463	0.31156
464	0.31156
465	0.311902
466	0.311902
467	0.311902
468	0.311902
469	0.311902
470	0.311902
471	0.311902
472	0.311902
473	0.311902
474	0.311902
475	0.311902
476	0.311902
477	0.311902
478	0.311902
479	0.311902
480	0.311902
481	0.312244
482	0.312244
483	0.312244
484	0.312244
485	0.312244
486	0.312244
487	0.312244
488	0.312244
489	0.312244
490	0.312244
491	0.312244
492	0.312244
493	0.312244
494	0.312244
495	0.312244
496	0.312244
497	0.312244
498	0.312244
499	0.312244
500	0.312244
501	0.312244
502	0.312244
503	0.312244
504	0.312244
505	0.312244
506	0.312244
507	0.312244
508	0.312244
509	0.312244
510	0.312244
511	0.312244
512	0.312244
513	0.312244
514	0.312244
515	0.312244
516	0.312244
517	0.312244
518	0.312244
519	0.312244
520	0.312244
521	0.312244
522	0.312244
523	0.312244
524	0.312244
525	0.312244
526	0.312244
527	0.312244
528	0.312244
529	0.312244
530	0.312244
531	0.312244
532	0.312244
533	0.312244
534	0.312244
535	0.312244
536	0.312244
537	0.312244
538	0.312244
539	0.312244
540	0.312244
541	0.312244
542	0.312244
543	0.340971
544	0.340971
545	0.340971
546	0.340971
547	0.340971
548	0.340971
549	0.340971
550	0.340971
551	0.340971
552	0.340971
553	0.340971
554	0.340971
555	0.340971
556	0.340971
557	0.340971
558	0.340971
559	0.344049
560	0.344049
561	0.344049
562	0.344049
563	0.344049
564	0.344049
565	0.344049
566	0.344049
567	0.344049
568	0.344049
569	0.344049
570	0.344049
571	0.344049
572	0.374145
573	0.374145
574	0.374145
575	0.374145
576	0.374145
577	0.374145
578	0.374145
579	0.374145
580	0.374145
581	0.374145
582	0.374145
583	0.374145
584	0.374145
585	0.374145
586	0.374145
587	0.374145
588	0.374145
589	0.374145
590	0.374145
591	0.374145
592	0.374145
593	0.374145
594	0.374145
595	0.374145
596	0.374145
597	0.408687
598	0.425787
599	0.425787
600	0.425787
601	0.438098
602	0.44699
603	0.44699
604	0.4487
605	0.4487
606	0.4487
607	0.4487
608	0.4487
609	0.4487
610	0.4487
611	0.463406
612	0.463406
613	0.46751
614	0.46922
615	0.46922
616	0.46922
617	0.47093
618	0.471614
619	0.471614
620	0.471614
621	0.471614
622	0.471614
623	0.471614
624	0.471614
625	0.471614
626	0.471614
627	0.471614
628	0.471956
629	0.471956
630	0.471956
631	0.471956
632	0.472298
633	0.472298
634	0.472298
635	0.472298
636	0.472298
637	0.472298
638	0.472298
639	0.472298
640	0.472298
641	0.472298
642	0.472298
643	0.472298
644	0.472298
645	0.472298
646	0.472298
647	0.472298
648	0.472298
649	0.472298
650	0.472298
651	0.499658
652	0.499658
653	0.511628
654	0.518126
655	0.518126
656	0.518126
657	0.518126
658	0.518126
659	0.52394
660	0.52394
661	0.526676
662	0.527702
663	0.527702
664	0.527702
665	0.528044
666	0.528044
667	0.528044
668	0.528044
669	0.528044
670	0.528044
671	0.528044
672	0.528044
673	0.528044
674	0.528044
675	0.52907
676	0.52907
677	0.529412
678	0.529412
679	0.529412
680	0.529412
681	0.529412
682	0.529412
683	0.529412
684	0.529412
685	0.529412
686	0.529412
687	0.529412
688	0.583105
689	0.583105
690	0.583105
691	0.583105
692	0.583105
693	0.583105
694	0.583105
695	0.583105
696	0.583105
697	0.583105
698	0.583105
699	0.583105
700	0.583105
701	0.583105
702	0.583105
703	0.583105
704	0.583105
705	0.583105
706	0.583105
707	0.583105
708	0.653899
709	0.653899
710	0.653899
711	0.653899
712	0.691176
713	0.691176
714	0.691176
715	0.691176
716	0.691176
717	0.691176
718	0.691176
719	0.691176
720	0.706566
721	0.706566
722	0.706566
723	0.706566
724	0.706566
725	0.706566
726	0.706566
727	0.706566
728	0.706566
729	0.706566
730	0.706566
731	0.706566
732	0.706566
733	0.706566
734	0.706566
735	0.706566
736	0.713406
737	0.713406
738	0.713406
739	0.713406
740	0.713406
741	0.713406
742	0.713406
743	0.713406
744	0.713406
745	0.713406
746	0.713406
747	0.713406
748	0.713406
749	0.713406
750	0.713406
751	0.713406
752	0.713406
753	0.713406
754	0.713406
755	0.713406
756	0.713406
757	0.713406
758	0.713406
759	0.713406
760	0.713406
761	0.713406
762	0.713406
763	0.713406
764	0.713406
765	0.713406
766	0.713406
767	0.713406
768	0.713406
769	0.713406
770	0.713406
771	0.713406
772	0.713406
773	0.713406
774	0.713406
775	0.720588
776	0.720588
777	0.720588
778	0.720588
779	0.721614
780	0.721614
781	0.721614
782	0.721614
783	0.721614
784	0.721614
785	0.721614
786	0.721614
787	0.722298
788	0.722298
789	0.722982
790	0.722982
791	0.722982
792	0.776334
793	0.776334
794	0.776334
795	0.776334
796	0.776334
797	0.776334
798	0.776334
799	0.776334
800	0.776334
801	0.776334
802	0.850205
803	0.850205
804	0.882011
805	0.882011
806	0.882011
807	0.882011
808	0.898769
809	0.898769
810	0.898769
811	0.898769
812	0.898769
813	0.898769
814	0.898769
815	0.898769
816	0.907661
817	0.907661
818	0.907661
819	0.907661
820	0.907661
821	0.907661
822	0.907661
823	0.907661
824	0.907661
825	0.907661
826	0.907661
827	0.907661
828	0.907661
829	0.907661
830	0.907661
831	0.907661
832	0.907661
833	0.945964
834	0.963748
835	0.963748
836	0.973324
837	0.973324
838	0.973324
839	0.973324
840	0.979138
841	0.979138
842	0.979138
843	0.979138
844	0.979138
845	0.979138
846	0.979138
847	0.979138
848	0.980848
849	0.980848
850	0.980848
851	0.980848
852	0.980848
853	0.980848
854	0.980848
855	0.980848
856	0.980848
857	0.980848
858	0.980848
859	0.980848
860	0.980848
861	0.980848
862	0.980848
863	0.980848
864	0.98119
865	0.98119
866	0.98119
867	0.98119
868	0.98119
869	0.98119
870	0.98119
871	0.98119
872	0.98119
873	0.98119
874	0.98119
875	0.98119
876	0.98119
877	0.98119
878	0.98119
879	0.98119
880	0.98119
881	0.98119
882	0.98119
883	0.98119
884	0.98119
885	0.98119
886	0.98119
887	0.98119
888	0.98119
889	0.98119
890	0.98119
891	0.98119
892	0.98119
893	0.98119
894	0.98119
895	0.98119
896	0.98119
897	0.988714
898	0.995554
899	0.995554
900	0.997948
901	0.997948
902	0.997948
903	0.997948
904	0.998632
905	0.998632
906	0.998632
907	0.998632
908	0.998632
909	0.998632
910	0.998632
911	0.998632
912	0.999316
913	0.999316
914	0.999316
915	0.999316
916	0.999316
917	0.999316
918	0.999316
919	0.999316
920	0.999316
921	0.999316
922	0.999316
923	0.999316
924	0.999316
925	0.999316
926	0.999316
927	0.999316
928	0.999658
929	0.999658
930	0.999658
931	0.999658
932	0.999658
933	0.999658
934	0.999658
935	0.999658
936	0.999658
937	0.999658
938	0.999658
939	0.999658
940	0.999658
941	0.999658
942	0.999658
943	0.999658
944	0.999658
945	0.999658
946	0.999658
947	0.999658
948	0.999658
949	0.999658
950	0.999658
951	0.999658
952	0.999658
953	0.999658
954	0.999658
955	0.999658
956	0.999658
957	0.999658
958	0.999658
959	0.999658
960	1
961	1
962	1
963	1
964	1
965	1
966	1
967	1
968	1
969	1
970	1
971	1
972	1
973	1
974	1
975	1
976	1
977	1
978	1
979	1
980	1
981	1
982	1
983	1
984	1
985	1
986	1
987	1
988	1
989	1
990	1
991	1
992	1
993	1
994	1
995	1
996	1
997	1
998	1
999	1
1000	1
1001	1
1002	1
1003	1
1004	1
1005	1
1006	1
1007	1
1008	1
1009	1
1010	1
1011	1
1012	1
1013	1
1014	1
1015	1
1016	1
1017	1
1018	1
1019	1
1020	1
1021	1
1022	1
1023	1
};
\addlegendentry{$E_b/N_0 = 3.0$ dB}

\end{axis}
\end{tikzpicture}

%% file: PSCF-iter.tikz
\begin{tikzpicture}
  \pgfplotsset{
    label style = {font=\fontsize{9pt}{7.2}\selectfont},
    tick label style = {font=\fontsize{7pt}{7.2}\selectfont}
  }

\begin{axis}[
	scale = 1,
    xlabel={$E_b/N_0$ [\text{dB}]}, xlabel style={yshift=0.4em},
    ylabel={Norm. Av. Computational Complexity}, ylabel style={yshift=-0.75em},
    grid=both,
    ymajorgrids=true,
    xmajorgrids=true,
    grid style=dashed,
    width=1\columnwidth, height=6cm,
    thick,
    mark size=3,
    legend cell align=left,
    legend columns=1,
]

\addplot[
    color=blue,
    mark=o,
    thick,
    mark size=3,
]
table {

1.00 4.26
1.20 3.59
1.40 2.84
1.60 2.17
1.80 1.67
2.00 1.34
2.20 1.15
2.40 1.06
2.60 1.02
2.80 1.01
3.00 1.00
3.50 1.00
};
\addlegendentry{SC-Flip}

\addplot[
    color=red,
    mark=x,
    thick,
    mark size=3,
]
table {
1.00 1.56
1.20 1.53
1.40 1.43
1.60 1.30
1.80 1.19
2.00 1.10
2.50 1.02
3.00 1.00
3.50 1.00
};
\addlegendentry{PSCF($P=2$)}

\addplot[
    color=green!50!black,
    mark=triangle,
    thick,
    mark size=3,
]
table {
1.00 0.857
1.20 0.920
1.40 0.965
1.60 0.993
1.80 1.00
2.00 1.01
2.50 1.01
3.00 1.00
3.50 1.00
};
\addlegendentry{PSCF($P=4$)}

\addplot[
    color=black,
    dashed,
    thick,
    mark size=3,
]
table {
1.0 1
3.5 1
};
\addlegendentry{SC}

\addplot[
    color=black,
    dotted,
    thick,
    mark size=3,
]
table {
1.0 2
3.5 2
};
\addlegendentry{SC-List ($L=2$)}

\addplot[
    color=black,
    dash dot,
    thick,
    mark size=3,
]
table {
1.0 4
3.5 4
};
\addlegendentry{SC-List ($L=4$)}

\end{axis}
\end{tikzpicture}

%% file: PSCF-perf-P2.tikz
\begin{tikzpicture}[spy using outlines=
	{rectangle, magnification=2, connect spies}]

\begin{axis}[
	scale = 1,
    ymode=log,
    y label style={at={(axis description cs:-0.05,.5)}},
    xlabel={$E_b/N_0$ [\text{dB}]}, xlabel style={yshift=0.4em},
    ylabel={FER}, ylabel style={yshift=-0.75em},
    grid=both,
    ymajorgrids=true,
    xmajorgrids=true,
    grid style=dashed,
    width=1\columnwidth, height=5.65cm,
    thick,
    mark size=3,
    legend style={
      anchor={center},
      cells={anchor=west},
      column sep= 2mm,
      font=\fontsize{7pt}{7.2}\selectfont,
    },
    legend to name=PSCF-perf,
    legend columns=2,
]

\addplot[
    color=black,
    mark=otimes,
    thick,
    mark size=3,
]
table {
1.0 7.28900e-01
1.5 3.36900e-01
2.0 8.44000e-02
2.5 1.24000e-02
3.0 1.18618e-03
3.5 1.08427e-04
4.0 1.20714e-05
};
\addlegendentry{SC}

\addplot[
    color=blue,
    mark=square,
    thick,
    mark size=3,
]
table {
1.00 7.39400e-01
1.50 3.09630e-01
2.00 5.89700e-02
2.50 4.86000e-03
3.00 2.08204e-04
3.50 6.23019e-06
};
\addlegendentry{SC-Flip}

\addplot[
    color=red,
    mark=triangle,
    thick,
    mark size=3,
]
table {
1.00 6.20240e-1
1.50 2.01990e-1
2.00 2.93400e-2
2.50 2.52000e-3
3.00 1.37574e-4
3.50 9.26778e-6
};
\addlegendentry{PSCF ($P = 2$, $M=1.0$dB)}

\addplot[
    color=black!60!green,
    mark=o,
    thick,
    mark size=3,
]
table {
1.00 6.14990e-1
1.50 1.97940e-1
2.00 2.76000e-2
2.50 2.11000e-3
3.00 1.30032e-4
3.50 9.88524e-6
};
\addlegendentry{PSCF ($P = 2$, $M=1.5$dB)}

\addplot[
    color=black,
    dashed,
    thick,
    mark size=3,
]
table {
1.00 6.57050e-01
1.50 2.18650e-01
2.00 2.87400e-02
2.50 1.39000e-03
3.00 6.78073e-05
3.50 3.10025e-06
};
\addlegendentry{SC-Oracle}

\addplot[
    color=orange!50!black,
    mark=diamond,
    thick,
    mark size=3,
]
table {
1 0.649 
1.5 0.2276 
2 0.0379
2.5 0.00322248
3 0.00014975 
3.5 7.96397e-06
};
\addlegendentry{SC-List ($L=2$)}


\coordinate (spypoint) at (axis cs:2.5,2e-3);
\coordinate (magnifyglass) at (axis cs:1.35,5e-5);

\end{axis}
\spy [magenta, height=2.2cm, width=2.2cm] on (spypoint)
   in node[fill=white] at (magnifyglass);
\end{tikzpicture}


%% file: PSCF-perf-P4.tikz
\begin{tikzpicture}[spy using outlines=
	{rectangle, magnification=2, connect spies}]

\begin{axis}[
	scale = 1,
    ymode=log,
    y label style={at={(axis description cs:-0.05,.5)}},
    xlabel={$E_b/N_0$ [\text{dB}]}, xlabel style={yshift=0.4em},
    ylabel={FER}, ylabel style={yshift=-0.75em},
    grid=both,
    ymajorgrids=true,
    xmajorgrids=true,
    grid style=dashed,
    width=1\columnwidth, height=5.65cm,
    thick,
    mark size=3,
    legend style={
      anchor={center},
      cells={anchor=west},
      column sep= 2mm,
      font=\fontsize{7pt}{7.2}\selectfont,
    },
    legend to name=PSCF-perf,
    legend columns=2,
]

\addplot[
    color=black,
    mark=otimes,
    thick,
    mark size=3,
]
table {
1.0 7.28900e-01
1.5 3.36900e-01
2.0 8.44000e-02
2.5 1.24000e-02
3.0 1.18618e-03
3.5 1.08427e-04
4.0 1.20714e-05
};
\addlegendentry{SC}

\addplot[
    color=blue,
    mark=square,
    thick,
    mark size=3,
]
table {
1.00 7.39400e-01
1.50 3.09630e-01
2.00 5.89700e-02
2.50 4.86000e-03
3.00 2.08204e-04
3.50 6.23019e-06
};
\addlegendentry{SC-Flip}

\addplot[
    color=red,
    mark=triangle,
    thick,
    mark size=3,
]
table {
1.00 6.19690e-01
1.50 2.27640e-01
2.00 4.31800e-02
2.50 5.01000e-03
3.00 5.70000e-04
3.50 4.27072e-05
};
\addlegendentry{PSCF ($P = 4$, $M=1.0$dB)}

\addplot[
    color=black!60!green,
    mark=o,
    thick,
    mark size=3,
]
table {
1.00 6.23250e-01
1.50 2.25070e-01
2.00 4.29900e-02
2.50 5.53000e-03
3.00 4.49749e-04
3.50 4.87987e-05
};
\addlegendentry{PSCF ($P = 4$, $M=1.5$dB)}

\addplot[
    color=black,
    dashed,
    thick,
    mark size=3,
]
table {
1.00 6.57050e-01
1.50 2.18650e-01
2.00 2.87400e-02
2.50 1.39000e-03
3.00 6.78073e-05
3.50 3.10025e-06
};
\addlegendentry{SC-Oracle}

\addplot[
    color=orange!50!black,
    mark=diamond,
    thick,
    mark size=3,
]
table {
1 0.649 
1.5 0.2276 
2 0.0379
2.5 0.00322248
3 0.00014975 
3.5 7.96397e-06
};
\addlegendentry{SC-List ($L=2$)}


\coordinate (spypoint) at (axis cs:2.5,2e-3);
\coordinate (magnifyglass) at (axis cs:1.35,5e-5);

\end{axis}
\spy [magenta, height=2.2cm, width=2.2cm] on (spypoint)
   in node[fill=white] at (magnifyglass);
\end{tikzpicture}
